\documentclass[]{pasj02} 
\usepackage[switch,mathlines]{lineno} 

\jyear{2024}
\Received{}
\Accepted{}

\usepackage{graphicx}
\usepackage{url}
\usepackage{rotating}

\begin{document} 

\title{ 
Q-band Line Survey Observations toward a Carbon-chain-rich Clump in the Serpens South Region}

\author{Kotomi \textsc{Taniguchi}\altaffilmark{1}\altemailmark}
\altaffiltext{1}{National Astronomical Observatory of Japan (NAOJ), National Institutes of Natural Sciences (NINS), 2-21-1 Osawa, Mitaka, Tokyo, 181-8588, Japan}
\email{kotomi.taniguchi@nao.ac.jp}

\author{Fumitaka \textsc{Nakamura}\altaffilmark{1,2,3}}
\altaffiltext{2}{Graduate Institute for Advanced Studies, SOKENDAI, 2-21-1 Osawa, Mitaka, Tokyo 181-8588, Japan}
\altaffiltext{3}{Department of Astronomy, The University of Tokyo, Hongo, Tokyo 113-0033, Japan}

\author{Sheng-Yuan \textsc{Liu}\altaffilmark{4}}
\altaffiltext{4}{Institute of Astronomy and Astrophysics, Academia Sinica, 11F of Astronomy-Mathematics Building, National Taiwan University, No 1, Sec., Roosevelt Road, Taipei, 10617 Taiwan}

\author{Tomomi \textsc{Shimoikura}\altaffilmark{5}}
\altaffiltext{5}{Faculty of Social Information Studies, Otsuma Women’s University, Chiyoda-ku,Tokyo, 102-8357, Japan}

\author{Chau-Ching \textsc{Chiong}\altaffilmark{4}}

\author{Kazuhito \textsc{Dobashi}\altaffilmark{6}}
\altaffiltext{6}{Department of Astronomy and Earth Sciences, Tokyo Gakugei University, 4-1-1 Nukuikitamachi, Koganei, Tokyo 184-8501, Japan}

\author{Naomi \textsc{Hirano}\altaffilmark{4}}

\author{Yoshinori \textsc{Yonekura}\altaffilmark{7}}
\altaffiltext{7}{Center for Astronomy, Ibaraki University, 2-1-1 Bunkyo, Mito, Ibaraki 310-8512, Japan}

\author{Hideko \textsc{Nomura}\altaffilmark{1,2}}

\author{Atsushi \textsc{Nishimura}\altaffilmark{8}}
\altaffiltext{8}{Nobeyama Radio Observatory, National Astronomical Observatory of Japan (NAOJ), National Institutes of Natural Sciences (NINS), 462-2 Nobeyama, Minamimaki, Minamisaku, Nagano 384-1305, Japan}

\author{Hideo \textsc{Ogawa}\altaffilmark{9}} 
\altaffiltext{9}{Department of Physics, Graduate School of Science, Osaka Metropolitan University, 1-1 Gakuen-cho, Naka-ku, Sakai, Osaka 599-8531, Japan}

\author{Chen \textsc{Chien}\altaffilmark{4}}

\author{Chin-Ting \textsc{Ho}\altaffilmark{4}}

\author{Yuh-Jing \textsc{Hwang}\altaffilmark{4}} 

\author{You-Ting \textsc{Yeh}\altaffilmark{4,10}}  
\altaffiltext{10}{Institute of Astronomy, National Tsing Hua University, No. 101, Section 2, Kuang-Fu Road, Hsinchu 30013, Taiwan}

\author{Shih-Ping \textsc{Lai}\altaffilmark{10}}

\author{Yasunori \textsc{Fujii}\altaffilmark{1}}

\author{Yasumasa \textsc{Yamasaki}\altaffilmark{1}}

\author{Quang \textsc{Nguyen-Luong}\altaffilmark{11}}
\altaffiltext{11}{Department of Computer Science, Mathematics \& Environmental Science, American University of Paris PL111, 2 bis, passage Landrieu, 75007 Paris, France}

\author{Ryohei \textsc{Kawabe}\altaffilmark{1}}



\maketitle

\begin{abstract}
We have conducted Q-band (30 GHz -- 50 GHz) line survey observations toward a carbon-chain emission peak in the Serpens South cluster-forming region with the extended Q-band (eQ) receiver installed on the Nobeyama 45 m radio telescope.
Approximately 180 lines have been detected including tentative detection, and these lines are attributed to 52 molecules including isotopologues.
It has been found that this position is rich in carbon-chain species as much as Cyanopolyyne Peak in Taurus Molecular Cloud-1 (TMC-1 CP), suggesting chemical youth. 
Not only carbon-chain species, but several complex organic molecules (CH$_3$OH, CH$_3$CHO, HCCCHO, CH$_3$CN, and tentatively C$_2$H$_3$CN) have also been detected, which is similar to the chemical complexity found in evolved prestellar cores.
The HDCS/H$_2$CS ratio has been derived to be $11.3 \pm 0.5$ \%, and this value is similar to the prestellar core L1544.
The chemically young features that are similar to the less-dense starless core TMC-1 CP ($10^4$ cm$^{-3}$ -- $10^5$ cm$^{-3}$) and chemically evolved characters which resemble the dense prestellar core L1544 ($\sim 10^6$ cm$^{-3}$) mean that the clump including the observed position is a {\it {pre-cluster clump}} without any current star formation activity.
\end{abstract}

\section{Introduction} \label{sec:intro}

Line survey observations are powerful methods to investigate chemical compositions comprehensively.
Various types of species can be covered in different frequency bands.
In the Q band, rotational lines from carbon-chain molecules including long chains ($e.g.,$ C$_6$H, HC$_7$N) can be efficiently detected.

Carbon-chain molecules account for $\sim43$\% of more than 300 interstellar molecules detected so far \citep{2024Ap&SS.369...34T}. 
Such carbon-chain species have been known as early-type species because they are abundant in the young starless core stage \citep{1998ApJ...506..743B,1992ApJ...392..551S,2024ApJ...965..162T}. 
Hence, it is essential to derive their chemical compositions for understanding the initial chemical characteristics of star formation.

High-sensitivity line survey observations in low-frequency bands ($i.e.,$ below Q-band) have been extensively performed.
Two large line survey projects, GOTHAM with the Green Bank 100 m telescope \citep{2020ApJ...900L..10M} and QUIJOTE with the Yebes 40 m telescope \citep{2021A&A...652L...9C}, have focused on Cyanopolyyne Peak in Taurus Molecular Cloud-1 (hereafter TMC-1 CP), which is the best studied carbon-chain-rich starless core \citep{2004PASJ...56...69K}.
These projects achieved the detection of many new carbon-chain species including large hydrocarbon rings and benzene ring(s)
; $e.g.,$ benzonitrile \citep{2018Sci...359..202M}, cyanonaphthalene \citep{2021Sci...371.1265M}, $cyclic$ ($c$)-C$_5$H$_6$ \citep{2021A&A...649L..15C}, $c$-C$_9$H$_8$ \citep{2021A&A...649L..15C, 2021ApJ...913L..18B}, o-C$_6$H$_4$ \citep{2021A&A...652L...9C}.
\citet{2024Ap&SS.369...34T} also summarize and discuss these recent updates.

Until now, line survey observations of starless cores have been conducted mainly toward nearby quiescent low-mass star-forming regions such as the Taurus region. 
However, it has been found that our Sun was born as a member of a cluster \citep{2010ARA&A..48...47A}.
In addition, most stars are born in cluster regions \citep{2003ARA&A..41...57L,2018ApJ...855...45S}.
Thus, it is essential to derive the chemical compositions of starless clumps/cores in cluster regions for understanding the initial chemical compositions of most stars and the solar system.
These starless cores are expected to be in different physical conditions from those in the well-studied quiescent low-mass star-forming regions.
For example, heating and the UV radiation from nearby (proto) stars could affect the chemical compositions of starless cores; the heating effect could change the degree of the deuterium fractionation ($e.g.,$ \cite{2019A&A...631A..25J}) and the UV radiation could enhance abundances of carbon-chain species \citep{2016A&A...592L..11S,2020A&A...643A..60S}.
Therefore, line survey observations toward starless clumps/cores in cluster regions and comparisons of the results among different physical conditions ($e.g.,$ with external radiation) are important to evaluate the effects of other cluster members on the chemical compositions at the starless core stage. 

This paper reports the Q-band (30 GHz -- 50 GHz) line survey observations toward a position showing the peak intensity of carbon-chain emission in the Serpens South cluster-forming region ($d=436 \pm 9$ pc; \cite{2017ApJ...834..143O}) with the extended Q-band (eQ) receiver \citep{2022SPIE12190E..0MC, 2024PASJ...76..563N}.
The Serpens South region consists of several filaments and is located at the west side of the W40 H$_{\rm {II}}$ region. 
\citet{2014ApJ...791L..23N} proposed a scenario that filament-filament collisions triggered the cluster formation.
Furthermore, \citet{2019PASJ...71S...4S} conducted large-scale mapping observations and proposed that the expanding shell of W40 plays an essential role in triggering star formation in the Serpens South cluster-forming region.
The Serpens South cluster-forming region contains a lot of dense cores and young protostars at various evolutionary stages ($e.g.,$ \cite{2008ApJ...673L.151G,2015A&A...584A..91K,2018A&A...615A...9P}), which indicates ongoing active star formation.
The excitation temperature and line width tend to be large at the central clump, presumably due to protostellar outflow feedback and stellar radiation \citep{2013ApJ...778...34T}.
On the other hand, the northern clump shows starless core features; high abundances of N$_2$H$^+$ and low dust temperatures \citep{2013ApJ...778...34T}.
\citet{2013MNRAS.436.1513F} obtained the map of the HC$_7$N ($J=21-20$) line with the Green Bank 100 m telescope and found that HC$_7$N is seen primarily toward cold filamentary structures within the region without protostars, largely avoiding the dense gas associated with small protostellar groups and embedded protostars within the main central cluster.
However, it is not clear how much the stellar feedback from the W40 H$_{\rm {II}}$ region and the Serpens South cluster affect the northern starless clump.

This paper consists as follows.
Section \ref{sec:obs} describes our observations.
Results and spectral analyses are presented in sections \ref{sec:result} and \ref{sec:ana}, respectively.
We will compare the chemical features in our target position to those in other starless cores in section \ref{sec:dis}.
Finally, the main points of this paper are summarized in section \ref{sec:sum}.

\section{Observations} \label{sec:obs}

\subsection{Mapping observations}

\begin{table}
  \tbl{Line list for mapping observations}{%
  \begin{tabular}{llcc}
      \hline
      Species & Transition & Frequency (GHz) & ${\rm {E}_{up}}/k$ (K) \\ 
      \hline
      HC$_5$N & $13-12$ & 34.614387 & 11.6 \\
      CH$_3$OH & $4_{-1,4}-3_{0,3}$ $E$ & 36.169261 & 28.8 \\
      HC$_3$N & $4-3$ & 36.392326 & 4.4 \\
      CCS & $4_3-3_2$ & 45.379033 & 5.4 \\
      $c$-C$_3$H$_2$ & $2_{1,1}-2_{0,2}$ & 46.755621 & 8.7 \\
      CH$_3$OH & $1_{0,1}-0_{0,0}$ $A$ & 48.37246 & 2.3 \\
      CS & $1-0$ & 48.990978 & 8.7 \\
      \hline
    \end{tabular}}\label{tab:otfline}
\begin{tabnote}
\end{tabnote}
\end{table}

We conducted mapping observations to obtain maps of the Serpens South region to determine an observed position of the line survey.
The mapping observations were conducted with the on-the-fly (OTF) mode using the eQ receiver in September 2023. 
We covered four carbon-chain species (HC$_3$N, HC$_5$N, CCS, and $c$-C$_3$H$_2$), CS, and two CH$_3$OH lines.
Table \ref{tab:otfline} summarizes information of the target lines.
The SAM 45 FX-type digital correlator was used. 
The channel separation and bandwidth were 15.26 kHz and 63 MHz, respectively.
We checked the telescope pointing every 1.5 hr -- 2 hr by observing the SiO maser line ($J=1-0$) from IRC+00363 at $(\alpha, \delta)_{\rm J2000.0}=$ (\timeform{18h41m25s.31}, $-$\timeform{04D20'32".1}).
The eQ receiver was used for the pointing observations.
The pointing accuracy was within 5\arcsec.

\begin{figure}
 \begin{center}
  \includegraphics[bb=0 15 250 470, scale=0.65]{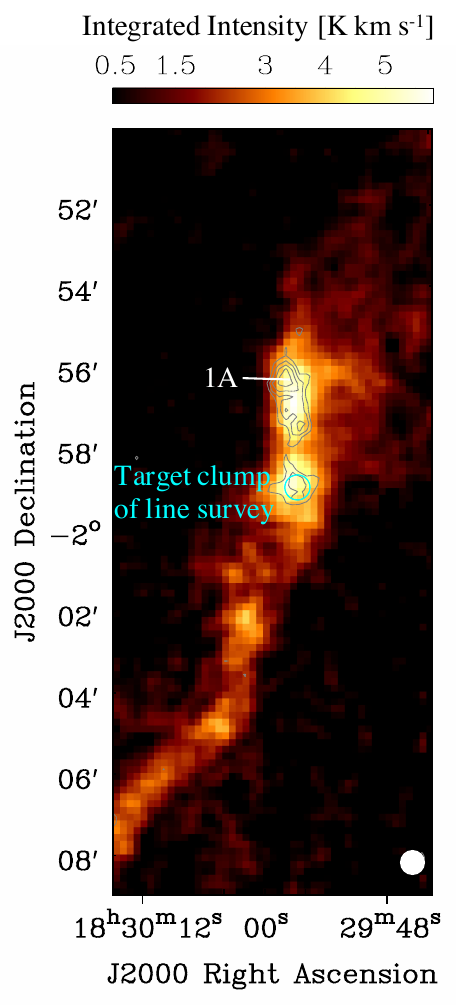} 
 \end{center}
\caption{Integrated-intensity maps of HC$_3$N overlaid by that of HC$_5$N (gray contours). The contour levels are 4,5,6,7 $\sigma$ and the noise level is 0.25 K\,km\,s$^{-1}$. The noise level of the HC$_3$N map is 0.4 K\,km\,s$^{-1}$. The cyan open circle indicates the position of the line survey observations at $(\alpha, \delta)_{\rm J2000.0}=$ (\timeform{18h29m56s.778}, $-$\timeform{1D58'56".7}). The size corresponds to the beam size of the Q-band (37\arcsec). The white-filled circle at the bottom right shows a typical beam size of the Q-band (37\arcsec). {Alt text: One two-dimensional molecular line emission map in the right ascension-declination coordinate. The unit of the map is Kelvin kilometer per second.}}\label{fig:map}
\end{figure}

We determined the target position of the line survey observations with the HC$_3$N map which shows the highest signal-to-noise (S/N) ratio.
Figure \ref{fig:map} shows integrated-intensity maps of HC$_3$N (color) overlaid by the HC$_5$N emission (gray contours).
The cyan open circle indicates our target position of the line survey observations.
The coordinate is $(\alpha, \delta)_{\rm J2000.0}=$ (\timeform{18h29m56s.778}, $-$\timeform{1D58'56".7}).
This position shows the highest peak intensity but narrow line widths compared to the northern one, which is referred to as Serpens South 1A \citep{2013MNRAS.436.1513F}.
The high peak intensity is advantageous for line detection.
Integrated-intensity maps of all of the molecular lines are presented in figure \ref{fig:map_all} in Appendix \ref{sec:a1}.
The systemic velocity ($V_{\rm {sys}}$) at this position is 7.5 km\,s$^{-1}$, which is determined from the C$^{18}$O ($J=1-0$) line data \citep{2019PASJ...71S...4S}\footnote{The data are available from the NRO Star Formation Project \citep{2019PASJ...71S...3N} Archive in Japanese Virtual Observatory (JVO; \url{http://jvo.nao.ac.jp/portal/nobeyama/sfp.do}).}.
The mass and virial ratio of the clump including the target position are 193 M$_{\Sol}$ and 0.10, respectively \citep{2013ApJ...778...34T}.
The mass is close to that of the central protocluster clump (232 M$_{\Sol}$), and thus the clump may eventually form a cluster.

\subsection{Line survey observations}

We carried out line survey observations between February and April 2024.
The eQ receiver was used for the observations.
The position-switching mode was employed.
The main beam efficiencies ($\eta_{\rm {MB}}$) and beam sizes (HPBW) are 75\% and 38.8\arcsec at 31 GHz, and 73\% and 36.6\arcsec at 43 GHz, respectively \citep{2024PASJ...76..563N}.
The beam size corresponds to $\approx 0.08$ pc at the source distance.
The off-source position was set at $-30$\arcmin \ in the right ascension direction from the on-source position.

We used the SAM 45 FX-type digital correlator.
We could not use four of 16 arrays of the backend SAM45 between February and April 2024 due to breakdown and only single-polarization data were obtained.
The bandwidth, channel separation, and frequency resolution were 63 MHz, 15.26 kHz, and 14.8 kHz, respectively.
The frequency resolution corresponds to a velocity resolution of $\sim0.1$ km\,s$^{-1}$.
The observations were conducted mostly in the 2SB mode (46 frequency setups).
We need three additional frequency setups covering 38.0 GHz -- 38.8 GHz with the single side-band (SSB) mode to cover the entire frequency range.
This frequency range corresponds to the edge of the 1st LO frequency and noise levels were higher than nearby other frequency ranges.
The system noise temperatures ($T_{\rm {sys}}$) were approximately 50 K -- 100 K in LSB and 80 K -- 270 K in USB, depending on the weather condition and frequency range.

We checked the telescope pointing every 1 hr -- 1.5 hr by observing the SiO maser line ($J=1-0$) from V1111 Oph at $(\alpha, \delta)_{\rm J2000.0}=$ (\timeform{18h37m19s.26}, $+$\timeform{10D25'42".2}) and IRC+00363. 
The eQ receiver was used for the pointing observations.
The pointing accuracy was less than 3\arcsec.

\section{Results} \label{sec:result}

\begin{figure*}
 \begin{center}
  \includegraphics[bb = 0 15 670 290, width=\textwidth]{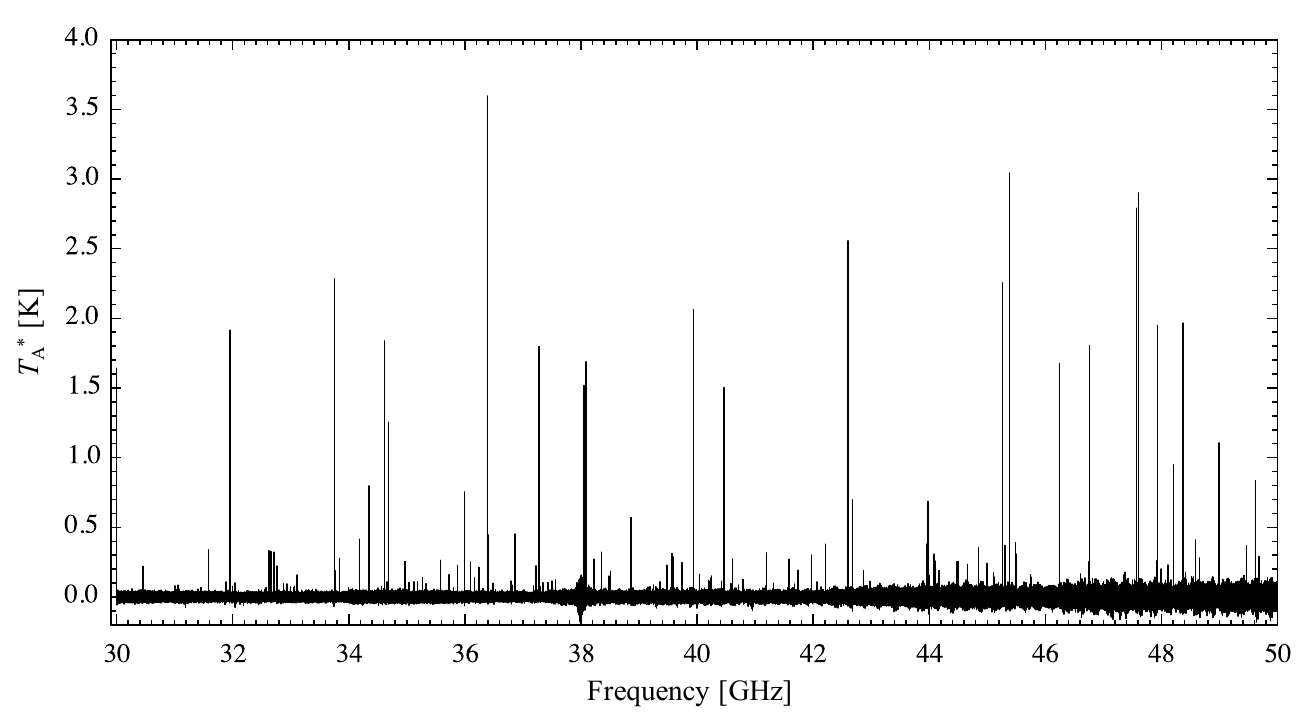} 
 \end{center}
\caption{Overview of the line survey data from 30 to 50 GHz. {Alt text: One graph. The vertical axis shows the antenna temperature in the unit of K and the horizontal axis shows frequency in the unit of GHz.}}\label{fig:allspec}
\end{figure*}

\begin{table*}
  \tbl{Detected molecules in the line survey\footnotemark[$*$]}{%
  \begin{tabular}{l|l}
      \hline
      Species & Isotopologues  \\  
      \hline
      \multicolumn{2}{l}{{\bf{Carbon-chain species}}} \\
      $l$-C$_3$H, $l$-C$_3$H$_{2}$, C$_4$H, $l$-C$_4$H$_{2}$, $l$-C$_5$H, C$_6$H, {\it {C$_6$H$^-$}}, $c$-C$_3$H$_{2}$, & $c$-C$_3$HD, $c$-H$^{13}$CCCH, DC$_3$N, H$^{13}$CCCN, HC$^{13}$CCN, \\
      HC$_3$N, HC$_5$N, HC$_7$N, HCCNC, HNCCC, C$_3$N, CCS, C$_3$S, & HCC$^{13}$CN, HCCC$^{15}$N, {\it {DC$_5$N}}, {\it {H$^{13}$CCCCCN}}, {\it {HC$^{13}$CCCCN}}, \\
        C$_3$O, CH$_3$CCH, CH$_3$C$_4$H, CH$_3$C$_3$N & {\it {HCC$^{13}$CCCN}}, {\it {HCCC$^{13}$CCN}}, {\it {HCCCC$^{13}$CN}}, CC$^{34}$S, C$_3$$^{34}$S \\
      \multicolumn{2}{l}{{\bf{COMs}}} \\
      CH$_3$OH, CH$_3$CHO, HCCCHO, CH$_3$CN & \\
      \multicolumn{2}{l}{{\bf{Other species}}} \\
       H$_2$CCO, {\it {C$_2$H$_3$CN}}, HNCO, SO, OCS, CS, HCS$^+$, H$_2$CS, {\it {HSCN}} & $^{13}$CS, C$^{34}$S, {\it {H$_2$C$\,^{34}$S}}, HDCS \\
      \hline
    \end{tabular}}\label{tab:mol}
\begin{tabnote}
\footnotemark[$*$] Species indicated in italics are tentative detection (S/N$\approx3$).  \\ 
\end{tabnote}
\end{table*}

\begin{center}
\footnotesize
\begin{longtable}{llccccc}
  \caption{Gaussian parameters of the detected lines}\label{tab:line}
  \hline              
  Species & Transition & Rest Frequency & $E_{\rm {up}}$ & $T_{\rm {A}}^*$ & FWHM     & $\Delta V_{\rm {LSR}}$  \\ 
          &            &  (GHz)         & (K) & (K)             & (km\,s$^{-1}$)      & (km\,s$^{-1}$) \\
\endfirsthead
  \hline
  Species & Transition & Rest Frequency & $E_{\rm {up}}$ & $T_{\rm {A}}^*$ & FWHM                & $\Delta V_{\rm {LSR}}$  \\ 
          &            &  (GHz)         & (K) & (K)             & (km\,s$^{-1}$)      & (km\,s$^{-1}$) \\ \hline \\
\endhead
  \hline
\endfoot
  \hline
\endlastfoot
  \hline
$l$-C$_3$H	&	$J=3/2-1/2$, $\Omega=1/2$, $F= 1- 1$, $l=f$	&	32.617016	&	1.6	&	0.050 (6) &	0.64 (9) & 0.27 (4) \\
$l$-C$_3$H	&	$J=3/2-1/2$, $\Omega=1/2$, $F= 2- 1$, $l=f$	&	32.627297	&	1.6	&	0.319 (7) &	0.55 (1) & 0.083 (6) \\
$l$-C$_3$H	&	$J=3/2-1/2$, $\Omega=1/2$, $F= 1- 0$, $l=f$	&	32.634389	&	1.6	&	0.121 (6) &	0.67 (4) &	0.06 (1) \\
$l$-C$_3$H	&	$J=3/2-1/2$, $\Omega=1/2$, $F= 2- 1$, $l=e$	&	32.660645	&	1.6	&	0.32 (1) &	0.56 (2) &	0.157 (9) \\
$l$-C$_3$H	&	$J=3/2-1/2$, $\Omega=1/2$, $F= 1- 0$, $l=e$	&	32.663361	&	1.6	&	0.14 (2) & 0.49 (8) & 0.17 (3) \\
$l$-C$_3$H	&	$J=3/2-1/2$, $\Omega=1/2$, $F= 1- 1$, $l=e$	&	32.667668	&	1.6	&	0.085 (7) &	0.55 (5) & -0.34 (2) \\
$l$-C$_3$H$_2$	&	$2_{1,2}-1_{1,1}$	&	41.198335	&	16.3 & 0.30 (1) & 0.47 (1) & -0.030 (8) \\
$l$-C$_3$H$_2$	&	$2_{0,2}-1_{0,1}$	&	41.584676	&	3.0	& 0.27 (1) & 0.51 (2) & -0.016 (9) \\
$l$-C$_3$H$_2$	&	$2_{1,1}-1_{1,0}$	&	41.967671	&	16.4 & 0.300 (8) & 0.51 (1) & -0.057 (7) \\
C$_4$H	&	$N = 4 - 3$, $J = 9/2 - 7/2$, $F = 4 - 3$	&	38.049616	&	4.6	& 1.48 (2) &	0.50 (1) & 0.027 (6) \\
C$_4$H	&	$N = 4 - 3$, $J = 9/2 - 7/2$, $F = 5 - 4$	&	38.049691	&	4.6	& 1.23 (2) &	0.51 (1) & 0.620 (8) \\
C$_4$H	&	$N = 4 - 3$, $J = 7/2 - 5/2$, $F = 4 - 3$; &	38.088441 & & & & \\
 & $N = 4 - 3$, $J = 7/2 - 5/2$, $F = 3 - 2$	 & 38.088481	&	4.6	& 1.71 (1) & 0.695 (6) & 0.009 (2) \\
C$_4$H	&	$N = 4 - 3$, $J = 7/2 - 7/2$, $F = 3 - 3$	&	38.212637	&	4.6	& 0.08 (1) &	0.40 (7) & 0.03 (3) \\
C$_4$H	&	$N = 5 - 4$, $J = 11/2 - 9/2$, $F = 5 - 4$; &	47.566770 & & & & \\
 & $N = 5 - 4$, $J = 11/2 - 9/2$, $F = 6 - 5$	 & 47.566814	&	6.9	& 2.87 (1) & 0.675 (5) & -0.133 (2) \\
C$_4$H	&	$N = 5 - 4$, $J = 9/2 - 7/2$, $F = 5 - 4$; &	47.605490 & & & & \\
 & $N = 5 - 4$, $J = 9/2 - 7/2$, $F = 4 - 3$	 & 47.605502 	&	6.9	& 2.85 (2) & 0.559 (5) & -0.034 (2) \\
$l$-C$_4$H$_2$	& $4_{1,4}-3_{1,3}$	&	35.577008	&	17.8	& 0.259 (9) & 0.50 (2) & -0.09 (8) \\
$l$-C$_4$H$_2$	& $4_{0,4}-3_{0,3}$	&	35.727379	&	4.3	& 0.154 (8) & 0.59 (3) & -0.25 (1) \\
$l$-C$_4$H$_2$	& $4_{1,3}-3_{1,2}$	&	35.875775	&	17.8	& 0.236 (9) & 0.55 (2) & 0.03 (1) \\
$l$-C$_4$H$_2$	& $5_{1,5}-4_{1,4}$	&	44.471138	&	19.9	& 0.25 (1) & 0.48 (3) & -0.04 (1) \\
$l$-C$_4$H$_2$	& $5_{0,5}-4_{0,4}$	&	44.659015	&	6.4	& 0.25 (2) & 0.47 (4) & -0.24 (1) \\
$l$-C$_4$H$_2$	& $5_{1,4}-4_{1,3}$	&	44.844590	&	20.0	& 0.37 (1) & 0.51 (2) &	0.04 (1) \\
$l$-C$_5$H	&	$J=13/2-11/2$, $\Omega=1/2$, $F= 6- 5$, $l=e$	&	31.029099	&	5.5	&	0.054 (9) &	0.50 (9) & 0.01 (4) \\
$l$-C$_5$H	&	$J=13/2-11/2$, $\Omega=1/2$, $F= 7- 6$, $l=e$	&	31.028815	&	5.5	&	0.038 (7) &	0.67 (16) & 0.03 (7) \\
$l$-C$_5$H	&	$J=13/2-11/2$, $\Omega=1/2$, $F= 7- 6$, $l=f$	&	31.032782	&	5.5	&	0.052 (9) &	0.58 (11) & -0.01 (4) \\
$l$-C$_5$H	&	$J=15/2-13/2$, $\Omega=1/2$, $F= 8- 7$, $l=e$	&	35.802773	&	7.2	&	0.069 (9) &	0.47 (7) & 0.10 (3) \\
$l$-C$_5$H	&	$J=15/2-13/2$, $\Omega=1/2$, $F= 7- 6$, $l=e$	&	35.803010	&	7.2	&	0.05 (1) & 0.47 (11) & 0.04 (4) \\
$l$-C$_5$H	&	$J=15/2-13/2$, $\Omega=1/2$, $F= 8- 7$, $l=f$	&	35.806812	&	7.2	&	0.069 (11) & 0.44 (8) & 0.09 (3) \\
$l$-C$_5$H	&	$J=15/2-13/2$, $\Omega=1/2$, $F= 7- 6$, $l=f$	&	35.807064	&	7.2	&	0.055 (8) &	0.70 (12) &	0.04 (5) \\
$l$-C$_5$H	&	$J=17/2-15/2$, $\Omega=1/2$, $F= 9- 8$, $l=e$	&	40.576710	&	9.2	&	0.10 (1) & 0.41 (5) & -0.13 (2) \\
$l$-C$_5$H	&	$J=17/2-15/2$, $\Omega=1/2$, $F= 8- 7$, $l=e$	&	40.576917	&	9.2	&	0.07 (1) & 0.51 (9) & -0.01 (3) \\
$l$-C$_5$H	&	$J=17/2-15/2$, $\Omega=1/2$, $F= 9- 8$, $l=f$	&	40.580839	&	9.2	&	0.09 (1) & 0.48 (7) & -0.11 (3) \\
$l$-C$_5$H	&	$J=17/2-15/2$, $\Omega=1/2$, $F= 8- 7$, $l=f$	&	40.581055	&	9.2	&	0.07 (1) & 0.55 (9) & -0.02 (3) \\
$l$-C$_5$H	&	$J=19/2-17/2$, $\Omega=1/2$, $F=10- 9$, $l=f$	&	45.354861	&	11.3	& 0.09 (1) & 0.39 (8) &	0.05 (3) \\
$l$-C$_5$H	&	$J=19/2-17/2$, $\Omega=1/2$, $F= 9- 8$, $l=f$	&	45.355054	&	11.3	& 0.09 (1) & 0.55 (10) & 0.10 (4) \\
C$_6$H	&	$J=23/2-21/2$, $\Omega=3/2$, $l=e$	&	31.881860	&	9.3	& 0.088 (5)	& 1.17 (8) & -0.03 (3) \\
C$_6$H	&	$J=23/2-21/2$, $\Omega=3/2$, $l=f$	&	31.885541	&	9.3	&	0.097 (5) &	1.10 (7) & -0.10 (3) \\
C$_6$H	&	$J=25/2-23/2$, $\Omega=3/2$, $l=e$	&	34.654037	&	11.0	& 0.102 (7) & 0.95 (7) & 0.08 (3) \\
C$_6$H	&	$J=25/2-23/2$, $\Omega=3/2$, $l=f$	&	34.658383	&	11.0	& 0.102 (6) & 0.97 (7) & 0.07 (3) \\
C$_6$H	&	$J=27/2-25/2$, $\Omega=3/2$, $l=e$	&	37.426192	&	12.8	& 0.089 (5)	&	0.97 (7) & -0.03 (3) \\
C$_6$H	&	$J=27/2-25/2$, $\Omega=3/2$, $l=f$	&	37.431255	&	12.8	& 0.099 (6) & 0.85 (6) & -0.07 (2) \\
C$_6$H	&	$J=29/2-27/2$, $\Omega=3/2$, $l=e$	&	40.198323	&	14.7	& 0.105 (6) &	0.95 (6) &	0.03 (2) \\
C$_6$H	&	$J=29/2-27/2$, $\Omega=3/2$, $l=f$	&	40.204157	&	14.7	& 0.124 (7) & 0.76 (5) & -0.01 (2) \\
C$_6$H	&	$J=31/2-29/2$, $\Omega=3/2$, $l=e$	&	42.970432	&	16.8	&	0.11 (1) &	0.73 (7) & -0.06 (3) \\
C$_6$H	&	$J=31/2-29/2$, $\Omega=3/2$, $l=f$	&	42.977089	&	16.8	& 0.102 (8) &	0.87 (8) & -0.10 (3) \\
C$_6$H	&	$J=33/2-31/2$, $\Omega=3/2$, $l=e$	&	45.742519	&	19.0	& 0.15 (2) &	0.77 (14) & -0.01 (6) \\
C$_6$H	&	$J=33/2-31/2$, $\Omega=3/2$, $l=f$	&	45.750052	&	19.0	& 0.13 (1) &	0.80 (7) & -0.01 (3) \\
C$_6$H	&	$J=35/2-33/2$, $\Omega=3/2$, $l=e$	&	48.514584	&	21.3	& 0.12 (1) &	0.58 (8) & 0.09 (3) \\
C$_6$H	&	$J=35/2-33/2$, $\Omega=3/2$, $l=f$	&	48.523044	&	21.3	& 0.14 (1) &	0.64 (8) & 0.01 (3) \\
C$_6$H$^-$	&	$15-14$	&	41.305453	&	15.9	& 0.04 (1) & 0.6 (2) & -0.1 (1) \\
$c$-C$_3$H$_2$	&	$3_{2,1}-3_{1,2}$	&	44.104777	&	18.2	& 0.25 (1) & 0.52 (3) &	-0.07 (1) \\
$c$-C$_3$H$_2$	&	$2_{1,1}-2_{0,2}$	&	46.755610	&	8.7	& 1.81 (1) & 0.576 (6) & -0.011 (2) \\
$c$-C$_3$HD	&	$2_{1,1}-2_{0,2}$	&	38.224437	&	7.6	&	0.24 (1) &	0.60 (3) & -0.04 (1) \\
$c$-C$_3$HD	&	$1_{1,1}-0_{0,0}$	&	49.615852	&	2.4	&	0.78 (2) & 0.64 (2) & 0.038 (9) \\
$c$-H$^{13}$CCCH	&	$2_{1,1}-2_{0,2}$	&	45.103867	&	8.5	&	0.17 (1) &	0.55 (5) &	0.17 (2) \\
HC$_3$N	&	$J= 4- 3$, $F= 4- 4$	&	36.390892	&	4.4	& 0.4 (3) & 0.4 (4) & 0.0 (1) \\
HC$_3$N	&	$J= 4- 3$, $F= 3- 2$	&	36.392326	&	4.4	& 2.32 (3) & 0.58 (1) & 0.733 (5) \\
	&		&		&		&	3.72 (3) & 0.759 (9) & -0.221 (3) \\
HC$_3$N	&	$J= 4- 3$, $F= 3- 3$	&	36.394169	&	4.4	& 0.4 (3) & 0.5 (4) & -0.1 (2) \\
HC$_3$N	&	$J= 5- 4$, $F= 5- 5$	&	45.488834	&	6.5	& 0.40 (2) & 0.49 (3) & 0.01 (1) \\
HC$_3$N	&	$J= 5- 4$, $F= 4- 4$	&	45.492106	&	6.5	& 0.31 (2) & 0.48 (5) & -0.11 (2) \\
DC$_3$N	&	$4-3$	&	33.772531	&	4.1	& 0.185 (7) & 0.77 (3) & -0.26 (1) \\
DC$_3$N	&	$5- 4$	&	42.215583	&	6.1	& 0.358 (7) & 0.72 (1) & 0.023 (7) \\
H$^{13}$CCCN	&	$4-3$	&	35.267403	&	4.2	& 0.124 (8) & 0.81 (6) & -0.18 (2) \\
H$^{13}$CCCN	&	$5-4$	&	44.084162	&	6.3	& 0.29 (1) & 0.58 (3) & -0.11 (1) \\
HC$^{13}$CCN	&	$4-3$	&	36.237949	&	4.3	& 0.15 (1) & 0.77 (7) & -0.14 (3) \\
HC$^{13}$CCN	&	$5-4$	&	45.297335	&	6.5	& 0.27 (2) & 0.62 (5) & 0.03 (2) \\
HCC$^{13}$CN	&	$4-3$	&	 36.241436	&	4.3	& 0.19 (1) & 0.73 (5) & -0.27 (2) \\
HCC$^{13}$CN	&	$5-4$	&	45.301707	&	6.5	& 0.35 (2) & 0.75 (6) & -0.02 (2) \\
HCCC$^{15}$N	&	$4-3$	&	35.333892	&	4.2	& 0.101 (9) & 0.48 (5) & 0.03 (2) \\
HCCC$^{15}$N	&	$5-4$	&	44.167268	&	6.4	& 0.17 (2) & 0.39 (5) & -0.02 (2) \\
HCCNC	&	$J= 4-3$, $F= 5-4$	&	39.742538	&	4.8	& 0.232 (9) & 0.52 (2) & -0.08 (1) \\
HCCNC	&	$J= 5-4$, $F= 6-5$	&	49.678064	&	7.2	& 0.27 (2) & 0.56 (5) & -0.03 (2) \\
HNCCC	&	$J= 4-3$, $F= 5-4$	&	37.3466	&	4.5	& 0.100 (7) & 0.54 (4) & 0.54 (1) \\
HNCCC	&	$J= 5-4$, $F= 6-5$	&	46.6831	&	6.7	& 0.14 (1) & 0.38 (5) & 0.23 (2) \\
HC$_5$N	&	$12-11$	&	31.951772	&	10.0 & 1.903 (8) & 0.585 (3) & -0.143 (1) \\
HC$_5$N	&	$13-12$	&	34.614387	&	11.6 & 1.834 (9) & 0.592 (3) & 0.057 (1) \\
HC$_5$N	&	$14-13$	&	37.276994	&	13.4 & 1.799 (7) & 0.606 (2) & 0.061 (1) \\
HC$_5$N	&	$15-14$	&	39.939591	&	15.3 & 2.03 (1) & 0.527 (3) & 0.075 (1) \\
HC$_5$N	&	$16-15$	&	42.602153	&	17.4 & 2.55 (1) & 0.505 (2) & -0.036 (1) \\
HC$_5$N	&	$17-16$	&	45.264720	&	19.6 & 2.25 (1) & 0.491 (3) & -0.057 (1) \\
HC$_5$N	&	$18-17$	&	47.927275	&	21.9 & 1.98 (8) & 0.51 (2) & 0.01 (1) \\
DC$_5$N	&	$15-14$	&	38.13138	&	14.6 & 0.07 (1) & 0.5 (1) & 0.14 (4) \\
DC$_5$N	&	$16-15$	&	40.673410	&	16.6 & 0.07 (1) & 0.28 (7) & -0.07 (3) \\
H$^{13}$CCCCCN	&	$13-12$	&	33.71334	&	11.3 & 0.05 (1) & 0.34 (8) & 0.07 (3) \\
H$^{13}$CCCCCN	&	$14-13$	&	36.306628	&	13.1 & 0.043 (8) & 0.5 (1) & 0.05 (5) \\
H$^{13}$CCCCCN	&	$15-14$	&	38.89991	&	14.9 & 0.07 (1) & 0.41 (9) & 0.26 (4) \\
H$^{13}$CCCCCN	&	$16-15$	&	41.493172	&	16.9 & 0.05 (1) & 0.29 (9) & 0.05 (4) \\
HC$^{13}$CCCCN	&	$13-12$	&	34.25968	&	11.5 & 0.05 (1) & 0.54 (13) & 0.02 (5) \\
HC$^{13}$CCCCN	&	$14-13$	&	36.894985	&	13.3 & 0.048 (9) & 0.37 (8) & -0.11 (3) \\
HCC$^{13}$CCCN	&	$16-15$	&	42.558032	&	17.4 & 0.07 (3) & 0.4 (2) & 0.12 (10) \\
HCCC$^{13}$CCN	&	$14-13$	&	37.24292	&	13.4 & 0.05 (1) & 0.30 (7) & -0.03 (3) \\
HCCCC$^{13}$CN	&	$13-12$	&	34.27245	&	11.5 & 0.05 (1) & 0.4 (1) &	-0.13 (4) \\
HCCCC$^{13}$CN	&	$14-13$	&	36.908725	&	13.3 & 0.039 (6) & 0.9 (1) & -0.14 (7) \\
HCCCC$^{13}$CN	&	$15-14$	&	39.54502	&	15.2 & 0.06 (1) & 0.34 (7) & -0.06 (3) \\
HC$_7$N	&	$27-26$	&	30.455745	&	20.5	&	0.203 (9) &	0.77 (4) &	0.09 (2) \\
HC$_7$N	&	$28-27$	&	31.583698	&	22.0	&	0.325 (9) &	0.52 (2) & -0.201 (7) \\
HC$_7$N	&	$29-28$	&	32.711682	&	23.5	&	0.321 (7) &	0.50 (1) &	0.121 (5) \\
HC$_7$N	&	$30-29$	&	33.839625	&	25.2	&	0.285 (8) &	0.52 (1) &	-0.194 (7) \\
HC$_7$N	&	$31-30$	&	34.967588	&	26.9	&	0.26 (1) &	0.55 (2) &	0.05 (1) \\
HC$_7$N	&	$32-31$	&	36.095534	&	28.6	&	0.256 (7) &	0.50 (1) & -0.108 (7) \\
HC$_7$N	&	$33-32$	&	37.223492	&	30.4	&	0.222 (8) &	0.55 (2) & -0.046 (9) \\
HC$_7$N	&	$34-33$	&	38.351449	&	32.2	&	0.33 (1) &	0.43 (1) & 0.089 (7) \\
HC$_7$N	&	$35-34$	&	39.479412	&	34.1	&	0.23 (1) &	0.50 (2) & 0.24 (1) \\
HC$_7$N	&	$36-35$	&	40.607327	&	36.1	&	0.26 (1) &	0.48 (2) & 0.013 (9) \\
HC$_7$N	&	$37-36$	&	41.735264	&	38.1	& 0.19 (1) & 0.38 (2) &	-0.01 (1) \\
HC$_7$N	&	$38-37$	&	42.863198	&	40.1	& 0.17 (1) & 0.50 (4)& -0.09 (1) \\
HC$_7$N	&	$39-38$	&	43.991129	&	42.2	&	0.11 (1) & 0.45 (8) & -0.02 (3) \\
HC$_7$N	&	$40-39$	&	45.119055	&	44.4	& 0.14 (6) & 0.4 (1) & 0.08 (8) \\
HC$_7$N	&	$41-40$	&	46.246978	&	46.6	& 0.13 (9) & 0.6 (5) & -0.1 (2) \\
HC$_7$N	&	$42-41$	&	47.374897	&	48.9	& 0.16 (2) & 0.51 (8) & -0.09 (3) \\
C$_3$N	&	$N= 4- 3$, $J=9/2-7/2$, $F=7/2-5/2$	& 39.571319	& 4.7	& 0.22(1) & 0.44 (4) & -0.60 (1) \\
C$_3$N	&	$N= 4- 3$, $J=9/2-7/2$, $F=9/2-7/2$; & 39.571326 & & & & \\
 & $N= 4-3$, $J=9/2-7/2$, $F=11/2-9/2$ & 39.571398 &	4.7	& 0.29 (11)	& 0.51 (3) & -0.014 (16) \\
C$_3$N	& $N= 4- 3$, $J=7/2-5/2$, $F=7/2-5/2$ &	39.590204 & & & & \\
 & $N= 4- 3$, $J=7/2-5/2$, $F=9/2-7/2$	 & 39.590212	&	4.8	& 0.28 (1) & 0.57 (2) & -0.05 (1) \\
C$_3$N	& $N= 5- 4$, $J=11/2-9/2$, $F=9/2-7/2$ & 49.466402 & & & & \\
 & $N= 5- 4$, $J=11/2-9/2$, $F=11/2-9/2$ & 49.466407 & & & & \\
 & $N= 5- 4$, $J=11/2-9/2$, $F=13/2-11/2$	& 49.466453	&	7.1	&	0.35 (1) & 0.65 (3) & -0.15 (1) \\
C$_3$O	&	$4-3$	&	38.486862	&	4.6	&	0.16 (1) & 0.41 (3) & -0.17 (1) \\
C$_3$O	&	$5-4$	&	48.108504	&	6.9	&	0.21 (9) & 0.5 (2) & 0.15 (10) \\
CCS	&	$N= 2- 1$, $J= 3- 2$	&	33.751370	&	3.2	& 2.286 (9) & 0.765 (3) & -0.115 (1) \\
CCS	&	$N= 3- 2$, $J= 3- 2$	&	38.86642	&	12.4	& 0.580 (8) & 0.71 (1) & 0.125 (5) \\
CCS	&	$N= 4- 3$, $J= 3- 2$	&	43.981019	&	12.9	& 0.67 (1) & 0.58 (1) & -0.088 (5) \\
CCS	&	$N= 3- 2$, $J= 4- 3$	&	45.379046	&	5.4	& 3.03 (1) & 0.743 (5) & 0.117 (2) \\
CC$^{34}$S	&	$N= 2- 1$, $J= 3- 2$	&	33.111839	&	3.2	& 0.153 (7) & 0.63 (3) &	0.09 (1) \\
CC$^{34}$S	&	$N= 3- 2$, $J= 4- 3$	&	44.497599	&	5.3	& 0.26 (1) & 0.57 (3) &	0.06 (1) \\
C$_3$S	&	$6-5$	&	34.684369	&	5.8	&	1.226 (9) &	0.677 (6) & 0.032 (2) \\
C$_3$S	&	$7-6$	&	40.465015	&	7.8	&	1.455 (9) &	0.644 (4) &	-0.066 (2) \\
C$_3$S	&	$8-7$	&	46.245624	&	10.0	& 1.67 (4) & 0.60 (1) & -0.019 (8)	\\
CCC$^{34}$S	&	$6-5$	&	33.844247	&	5.7	& 0.038 (6) & 0.81 (14) & 0.14 (6) \\
CCC$^{34}$S	&	$7-6$	&	39.484877	&	7.6	& 0.066 (9) & 0.62 (9)& 0.05 (4) \\
CCC$^{34}$S	&	$8-7$	&	45.125471	&	9.7	& 0.12 (1) & 0.44 (7) &	0.14 (3) \\
CH$_3$CCH	&	$2_1-1_1$	&	34.182760	&	9.7	& 0.30 (2) & 0.58 (5) & -0.04 (2) \\
CH$_3$CCH	&	$2_0-1_0$	&	34.183414	&	2.5	& 0.42 (2) & 0.53 (3) & -0.03 (1) \\
CH$_3$C$_4$H	&	$8_1-7_1$	&	32.571458	&	15.3	& 0.043 (7) & 0.80 (15) & 0.05 (6) \\
CH$_3$C$_4$H	&	$8_0-7_0$	&	32.571775	&	7.4	& 0.059 (6) & 0.78 (10) & 0.03 (4) \\
CH$_3$C$_4$H	&	$9_1-8_1$	&	36.642837	&	17.2	& 0.05 (1) & 0.31 (7) & 0.03 (3) \\
CH$_3$C$_4$H	&	$9_0-8_0$	&	36.643194	&	9.2	& 0.048 (8) & 0.52 (9) & 0.05 (4) \\
CH$_3$C$_4$H	&	$10_1-9_1$	&	40.714197	&	19.2	& 0.08 (1) & 0.39 (5) & -0.05 (2) \\
CH$_3$C$_4$H	&	$10_0-9_0$	&	40.714594	&	11.3	& 0.06 (1) & 0.42 (8) &	-0.04 (3) \\
CH$_3$C$_3$N	&   $8_1-7_1$   & 33.051300 & 14.6 & 0.036 (9) & 0.53 (16) & -0.01 (6) \\
CH$_3$C$_3$N	&   $8_0-7_0$   & 33.051619	& 7.1 & 0.06 (1) & 0.44 (8) & -0.15 (3) \\
CH$_3$C$_3$N	&   $9_1-8_1$   & 37.182656 & 16.4 & 0.064 (9) & 0.51 (8) & -0.12 (3) \\
CH$_3$C$_3$N	&   $9_0-8_0$   & 37.183014 & 8.9 &	0.078 (8) & 0.51 (6) & -0.05 (2) \\
CH$_3$C$_3$N	&   $10_1-9_1$  & 41.313991 & 18.4 & 0.08 (1) & 0.57 (8) & -0.16 (3) \\
CH$_3$C$_3$N	&   $10_0-9_0$  & 41.314389 & 10.9 & 0.11 (1) & 0.48 (5) & -0.18 (2) \\
CH$_3$C$_3$N	&   $11_1-10_1$ & 45.445304	& 20.6 & 0.09 (1) & 0.41 (8) & 0.04 (3) \\
CH$_3$C$_3$N	&   $11_0-10_0$ & 45.445742 & 13.1 &  0.06 (1) & 0.6 (1) & -0.02 (6) \\
CH$_3$OH	&	$4_{-1,4}-3_{0,3}$ $E$, $v_t=0$	&	36.169261	&	28.8 & 0.133 (6) & 0.70 (4) & 0.00 (1) \\
CH$_3$OH	&	$1_{0,1}-0_{0,0}$ $A$, $v_t=0$	&	48.37246	&	2.3	& 1.91 (2) & 0.81 (1) & -0.011 (4) \\
CH$_3$OH	&	$1_{0,1}-0_{0,0}$ $E$, $v_t=0$ &	48.376887	&	15.4 & 0.16 (17) & 0.2 (2) & -0.1 (1) \\
CH$_3$CHO	&	$2_{1,2}-1_{1,1}$ $A$, $v_t=0$	&	37.464204	&	5.0	& 0.045 (7) & 0.59 (10) &	0.08 (4) \\
CH$_3$CHO	&	$2_{1,2}-1_{1,1}$ $E$, $v_t=0$	&	37.686932	&	5.0	& 0.06 (1) & 0.53 (11) & 0.10 (5) \\
CH$_3$CHO	&	$2_{0,2}-1_{0,1}$ $E$, $v_t=0$	&	38.506035	&	2.9	& 0.174 (8) & 0.58 (3) & 0.01 (1) \\
CH$_3$CHO	&	$2_{0,2}-1_{0,1}$ $A$, $v_t=0$	&	38.512079	&	2.8	& 0.155 (7) & 0.59 (3) & 0.04 (1) \\
CH$_3$CHO	&	$2_{1,1}-1_{1,0}$ $E$, $v_t=0$	&	39.362537	&	5.2	& 0.093 (9) & 0.57 (6) & 0.06 (2) \\
CH$_3$CHO	&	$2_{1,1}-1_{1,0}$ $A$, $v_t=0$	&	39.594289	&	5.1	& 0.106 (9) & 0.60 (6) & -0.05 (2) \\
H$_2$CCO	&	$2_{1,2}-1_{1,1}$	&	40.039022	&	15.9 & 0.137 (7) & 0.62 (4) & 0.08 (1) \\
H$_2$CCO	&	$2_{0,2}-1_{0,1}$	&	40.41795	&	2.9	& 0.09 (1) & 0.70 (9) & 0.01 (4) \\
H$_2$CCO	&	$2_{1,1}-1_{1,0}$	&	40.793832	&	16.0 & 0.111 (6) & 0.72 (4) & -0.05 (1) \\
HCCCHO	&	$4_{0,4}-3_{0,3}$	&	37.290140	&	4.5	& 0.069 (6) & 0.67 (7) & -0.11 (3) \\
HCCCHO	&	$5_{0,5}-4_{0,4}$	&	46.602874	&	6.7	& 0.12 (1) & 0.6 (1) & -0.18 (4) \\
CH$_3$CN	&	$2_0-1_0$	&	36.795475	&	2.7	& 0.111 (8) & 0.62 (7) & -0.86 (3) \\
            &               &               &       & 0.067 (9) & 0.45 (9) & -0.07 (4) \\
CH$_3$CN	&	$2_1-1_1$	&	36.794765	&	9.8	& 0.059 (9) & 0.66 (12) & -2.26 (5) \\
C$_2$H$_3$CN	&	$5_{0,5}-4_{0,4}$	&	47.354648	&	6.8	& 0.16 (1) & 0.54 (7) &	-0.10 (3) \\
C$_2$H$_3$CN	&	$4_{1,4}-3_{1,3}$	&	37.018922	&	6.6	& 0.042 (4) & 1.44 (16) &	0.09 (7) \\
C$_2$H$_3$CN	&	$5_{1,5}-4_{1,4}$	&	46.266934	&	8.8	& 0.08 (1) & 0.87 (14) &	-0.00 (6) \\
HNCO	&	$2_{0,2}-1_{0,1}$	&	43.963040	& 3.2	& 0.35 (1) & 0.61 (2) & 0.20 (1) \\
SO & $J_N=1_0-0_1 $ & 30.001580 & 1.4	& 1.66 (1) & 1.108 (8) & 0.531 (3) \\
OCS	&	$3-2$	&	36.488812	&	3.5	& 0.087 (6) & 0.93 (7) & 0.10 (3) \\
OCS	&	$4-3$	&	48.651604	&	5.8	& 0.25 (2) & 0.68 (6) & -0.06 (2) \\
CS	&	$1-0$	&	48.990955	&	2.4	& 0.67 (2) & 0.67 (2) & -0.675 (8) \\
&		&		&		& 0.55 (1) & 2.26 (4) & 0.11 (3) \\
$^{13}$CS	&	$1-0$	&	46.247563	&	2.2	& 0.71 (7) & 0.87 (9) &	-0.02 (4) \\
C$^{34}$S	&	$1-0$	& 48.206941	&	2.3	& 0.967 (1) & 1.07 (1) & 0.001 (6) \\
HCS$^+$	&	$1-0$	&	42.674195	&	2.0	& 0.66 (1) & 0.71 (1) & 0.058 (6) \\
H$_2$CS	&	$1_{0,1}-0_{0,0}$	&	34.35143	&	1.6	& 0.771 (9) & 0.75 (1) & 0.050 (4) \\
H$_2$C$^{34}$S	&	$1_{0,1}-0_{0,0}$	&	33.7658	&	1.6	& 0.036 (6) & 0.83 (16) & 0.22 (7) \\
HDCS	&	$1_{0,1}-0_{0,0}$	&	31.002299	&	1.5	& 0.086 (7) & 0.84 (8) & 0.10 (3) \\
HSCN	&	$3_{0,3}-2_{0,2}$	&	34.408629	&	3.3	& 0.04 (1) & 0.5 (1) & -0.48 (5) \\
HSCN	&	$4_{0,4}-3_{0,3}$	&	45.877809	&	5.5	& 0.08 (1) & 1.4 (2) & 0.1 (1) \\
\hline
\multicolumn{7}{l}{$*$Numbers in parentheses are the standard deviation, expressed in units of the last significant digits.} \\ 
\end{longtable}
\end{center}

Figure \ref{fig:allspec} shows an overall spectrum of the line survey from 30 to 50 GHz.
The integration times for each frequency setup are between 45 min and 80 min.
The spectral noise levels were approximately 10 mK -- 20 mK depending on the frequency range.
We have higher system temperatures in higher frequencies.
These noise levels are comparable with the results of the line survey toward TMC-1 CP by \citet{2004PASJ...56...69K}.
Figures \ref{fig:s1} in Appendix \ref{sec:a2} show close-up spectra.

We identified the detected emission lines using the CASSIS software \citep{2015sf2a.conf..313V} with spectroscopic databases of JPL \citep{1998JQSRT..60..883P} and the Cologne Database for Molecular Spectroscopy (CDMS; \cite{2005JMoSt.742..215M,2016JMoSp.327...95E}). 
We have detected $\sim180$ lines from 52 molecular species including isotopologues.
Table \ref{tab:mol} summarizes the detected species.
We have detected almost all of the species that \citet{2004PASJ...56...69K} detected toward TMC-1 CP in the same frequency range, except for CCO, HC$_3$NH$^+$, and CH$_2$CN.
If their peak intensities are similar to those in TMC-1 CP (51 mK and 200 mK for HC$_3$NH$^+$ and CH$_2$CN), the HC$_3$NH$^+$ and CH$_2$CN lines would be detected since the noise levels of our data are $\sim10$ mK.
Thus, they are deficient in our target position compared to TMC-1 CP.
On the other hand, the non-detection of CCO (45.826725 GHz) may be caused by a high noise level around this frequency band, because its peak intensity in TMC-1 CP is only 27 mK \citep{2004PASJ...56...69K}.

We conducted Gaussian fitting of the detected lines, including tentative detection (S/N$\approx3$).
Table \ref{tab:line} summarizes the results of the Gaussian fitting\footnote{Numbers in parentheses in Table \ref{tab:line} are the standard deviation of the Gaussian fit, expressed in units of the last significant digits.}.
The column of $\Delta V_{\rm {LSR}}$ indicates shifts from the systemic velocity (7.5 km\,s$^{-1}$).
Three lines, CS ($1-0$), HC$_{3}$N ($J= 4- 3$, $F= 3- 2$), and CH$_3$CN ($J_K=2_0-1_0$), were fitted by two velocity components.
Some C$_4$H and C$_3$N lines are blended and we fitted these spectra as one line.
Figure \ref{fig:3spec} in Appendix \ref{sec:a2} show examples of the observed spectra overlaid by the Gaussian fitting results.

\section{Spectral Analyses} \label{sec:ana}

\begin{figure*}
 \begin{center}
  \includegraphics[bb = 0 10 900 190, width=\textwidth]{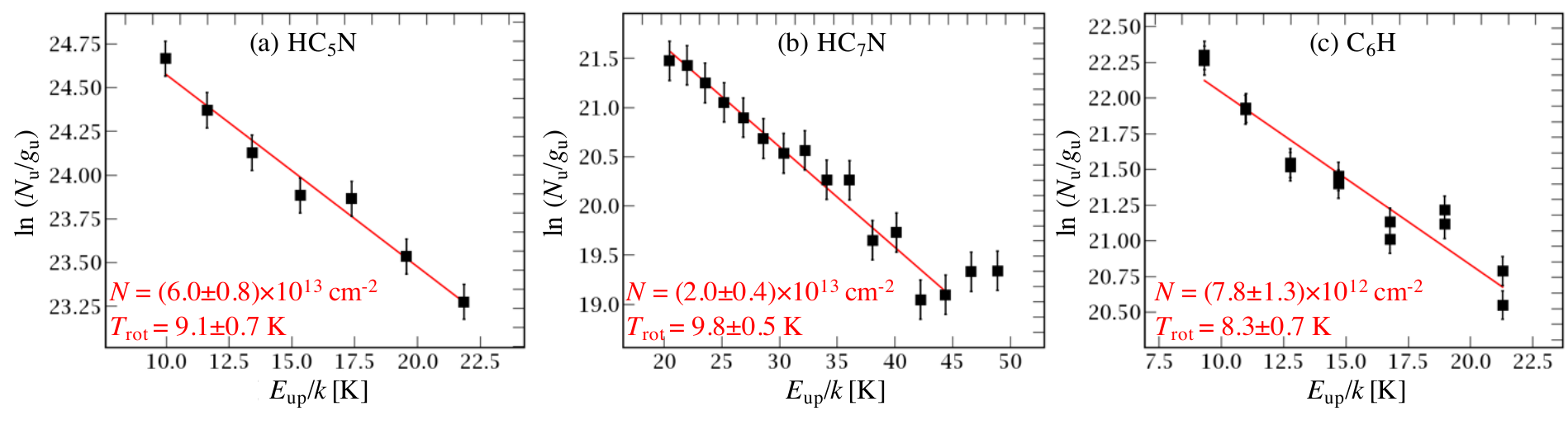} 
 \end{center}
\caption{Rotational diagram of (a) HC$_5$N, (b) HC$_7$N, and (c) C$_6$H. We analyzed HC$_7$N using only lines covered in LSB. The error bars include the Gaussian fitting and absolute calibration error (10\%). The beam-filling factor was included assuming that the emission size is 30\arcsec. The fitting results are indicated by red lines and derived values are shown in each panel. {Alt text: Three panels. The vertical axis shows ln($N_u$/$g_u$) with no unit and the horizontal axis shows the upper-state energies in the unit of Kelvin.}}\label{fig:RD}
\end{figure*}

We conducted spectral analyses with the rotational diagram method for carbon-chain species of which more than four lines with different upper-state energies have been detected.
We used the CASSIS software for the analyses.
We included the beam-filling factor assuming that the emission-region size is 30\arcsec, which was constrained by the integrated-intensity map of HC$_5$N (panel b of figure \ref{fig:map_all}).
The absolute calibration error of 10\%, a typical value of the chopper-wheel method, was taken into consideration.
Three species (HC$_5$N, HC$_7$N, and C$_6$H) were fitted.
In the case of HC$_7$N, two lines with the highest upper-state energies were excluded from the fitting because they show systematically higher values. 
Figure \ref{fig:RD} shows their rotational diagram.
The derived column densities and rotational temperatures are ($6.0\pm0.8$)$\times10^{13}$ cm$^{-2}$ and $9.1\pm0.7$ K for HC$_5$N, ($2.0\pm0.4$)$\times10^{13}$ cm$^{-2}$ and $9.8\pm0.5$ K for HC$_7$N, and ($7.8\pm1.3$)$\times10^{12}$ cm$^{-2}$ and $8.3\pm0.7$ K for C$_6$H, respectively.
These rotational temperatures ($T_{\rm {rot}}$) are higher than a typical value of carbon-chain species in TMC-1 CP (6.5 K; \cite{1992ApJ...392..551S,2016ApJ...817..147T}).


\begin{table*}
\begin{center}
  \tbl{Column densities and fractional abundances with respect to H$_2$}{%
  \begin{tabular}{lcc|lcc}
      \hline
      Species & $N$ (cm$^{-2}$)\footnotemark[$*$] & Abundance\footnotemark[$*$] & Species & $N$ (cm$^{-2}$)\footnotemark[$*$] & Abundance\footnotemark[$*$] \\ 
      \hline
$l$-C$_3$H	& $(2.6\pm0.1)\times10^{13}$ & $(3.2\pm0.1)\times10^{-10}$	&	C$_3$O	&	$(3.6\pm0.3)\times10^{12}$ & $(4.4\pm0.4)\times10^{-11}$ \\
$l$-C$_3$H$_2$	& $(4.4\pm0.2)\times10^{12}$ & $(5.4\pm0.3)\times10^{-11}$ & CCS & $(2.3\pm0.1)\times10^{14}$ & $(2.9\pm0.2)\times10^{-9}$ \\
C$_4$H & $(2.15\pm0.07)\times10^{14}$ & $(2.66\pm0.09)\times10^{-9}$ & CC$^{34}$S & $(6.6\pm0.4)\times10^{12}$ & $(8.1\pm0.5)\times10^{-11}$ \\
$l$-C$_4$H$_2$ & $(4.3\pm0.2)\times10^{12}$	& $(5.3\pm0.3)\times10^{-11}$ &C$_3$S &	$(2.6\pm0.1)\times10^{13}$ & $(3.3\pm0.1)\times10^{-10}$ \\
$l$-C$_5$H & $(2.7\pm0.1)\times10^{12}$	& $(3.4\pm0.1)\times10^{-11}$ & C$_3$$^{34}$S & $(1.1\pm0.1)\times10^{12}$ & $(1.4\pm0.1)\times10^{-11}$ \\
C$_6$H & $(6.0\pm0.8)\times10^{13}$	& $(7.4\pm0.9)\times10^{-10}$ &CH$_3$CCH & $(2.77\pm0.08)\times10^{14}$ & $(3.4\pm0.1)\times10^{-10}$ \\
C$_6$H$^-$ & $<2.7\times10^{11}$ & $<3.4\times10^{-12}$ & CH$_3$C$_4$H & $(1.84\pm0.09)\times10^{13}$ & $(2.3\pm0.1)\times10^{-10}$ \\
$c$-C$_3$H$_2$ & $(2.5\pm0.2)\times10^{14}$	& $(3.1\pm0.2)\times10^{-9}$ & CH$_3$C$_3$N & $(1.7\pm0.1)\times10^{12}$ & $(2.1\pm0.1)\times10^{-11}$	\\
$c$-C$_3$HD	&	$(2.0\pm0.1)\times10^{13}$	& $(2.4\pm0.2)\times10^{-10}$ & CH$_3$OH ($A$) & $(6.6\pm0.3)\times10^{14}$ & $(8.2\pm0.4)\times10^{-9}$ \\
$c$-H$^{13}$CCCH & $(8.3\pm0.8)\times10^{12}$ & $(1.0\pm0.1)\times10^{-10}$	& CH$_3$OH ($E$) &	$(3.1\pm0.2)\times10^{14}$ & $(3.9\pm0.2)\times10^{-9}$	\\
HC$_3$N	& $(2.23\pm0.09)\times10^{14}$	&$(2.8\pm0.1)\times10^{-9}$ & CH$_3$CHO	& $(2.8\pm0.1)\times10^{13}$ & $(3.5\pm0.1)\times10^{-10}$ \\
DC$_3$N	& $(3.8\pm0.2)\times10^{12}$ & $(4.9\pm0.2)\times10^{-11}$ & H$_2$CCO & $(3.9\pm0.1)\times10^{13}$ & $(4.8\pm0.2)\times10^{-10}$ \\
H$^{13}$CCCN & $(2.4\pm0.1)\times10^{12}$ & $(3.0\pm0.2)\times10^{-11}$	& HCCCHO & $(8.9\pm0.9)\times10^{12}$ & $(1.1\pm0.1)\times10^{-10}$ \\
HC$^{13}$CCN & $(2.5\pm0.1)\times10^{12}$ & $(3.3\pm0.1)\times10^{-11}$ & CH$_3$CN & $(2.9\pm0.3)\times10^{12}$	& $(3.6\pm0.3)\times10^{-11}$ \\
HCC$^{13}$CN & $(3.1\pm0.1)\times10^{12}$ & $(3.9\pm0.2)\times10^{-11}$	& C$_2$H$_3$CN	& $<4.2\times10^{12}$ & $<5.2\times10^{-11}$ \\
HCCC$^{15}$N & $(1.2\pm0.1)\times10^{12}$ & $(1.5\pm0.1)\times10^{-11}$ &	HNCO & $(1.63\pm0.07)\times10^{13}$	& $(2.02\pm0.09)\times10^{-10}$ \\
HCCNC &	$(3.5\pm0.2)\times10^{12}$ & $(4.5\pm0.2)\times10^{-11}$ & SO & $9.1^{+0.9}_{-1.2}\times10^{13}$ & $(1.1\pm0.1)\times10^{-9}$ \\
HNCCC &	$(4.9\pm0.4)\times10^{11}$ & $(6.2\pm0.5)\times10^{-12}$ & OCS & $(5.3\pm0.3)\times10^{13}$	& $(6.6\pm0.4)\times10^{-10}$ \\
HC$_5$N	& $(6.0\pm0.8)\times10^{13}$ & $(7.4\pm0.9)\times10^{-10}$ & CS (1st)$\dag$	& $(3.33\pm0.04)\times10^{13}$ & $(4.13\pm0.05)\times10^{-10}$ \\
DC$_5$N	& $<1.2\times10^{12}$ & $<1.5\times10^{-11}$ & CS (2nd) $\dag$ & $(2.46\pm0.03)\times10^{13}$ & $(3.04\pm0.04)\times10^{-10}$ \\
H$^{13}$CCCCCN & $<7.8\times10^{11}$ & $<9\times10^{-12}$ & $^{13}$CS	& $(2.4\pm0.1)\times10^{13}$ & $(2.9\pm0.2)\times10^{-10}$ \\
HC$^{13}$CCCCN & $<9.4\times10^{11}$ & $<1.2\times10^{-11}$ & C$^{34}$S	& $(3.6\pm0.2)\times10^{13}$ & $(4.4\pm0.3)\times10^{-10}$ \\
HCC$^{13}$CCCN & $<1.2\times10^{12}$ & $<1.5\times10^{-11}$ & HCS$^+$ & $(2.2\pm0.1)\times10^{13}$ & $(2.8\pm0.1)\times10^{-10}$ \\
HCCC$^{13}$CCN & $<7.9\times10^{11}$ & $<9.7\times10^{-12}$ & H$_2$CS &	$(2.12\pm0.09)\times10^{14}$ & $(2.6\pm0.1)\times10^{-9}$ \\
HCCCC$^{13}$CN & $<1.1\times10^{12}$ & $<1.3\times10^{-11}$ & H$_2$C$^{34}$S & $<1.0\times10^{13}$ & $<1.3\times10^{-10}$ \\
HC$_7$N	& $(2.0\pm0.4)\times10^{13}$ & $(2.5\pm0.4)\times10^{-10}$ & HDCS &	$(2.4\pm0.1)\times10^{13}$ & $(3.0\pm0.2)\times10^{-10}$ \\
C$_3$N & $(1.65\pm0.04)\times10^{13}$ & $(2.05\pm0.05)\times10^{-10}$ &	HSCN & $<2.1\times10^{12}$ & $<2.6\times10^{-11}$\\
      \hline
    \end{tabular}}\label{tab:column}
\begin{tabnote}
\footnotemark[$*$] We add ''$<$'' marks for the tentative detection species. \\ 
\footnotemark[$\dag$] The first and second components of the CS line are $\Delta\,V_{\rm {LSR}} \approx -0.68$ km\,s$^{-1}$ and 0.1 km\,s$^{-1}$, respectively. 
\end{tabnote}
\end{center}
\end{table*}

We analyzed spectra with the Markov Chain Monte Carlo (MCMC) method in the CASSIS software for the other species. 
We assumed the local thermodynamic equilibrium (LTE) conditions.
In the MCMC analyses, the column density ($N$), excitation temperature ($T_{\rm {ex}}$), line width (FWHM), velocity centroid ($V_{\rm {LSR}}$), and emission size are treated as free parameters. 
We fixed the excitation temperature at 10 K which was constrained by the rotational diagram of the three species (HC$_5$N, HC$_7$N, and C$_6$H).
We fixed the emission size at 30\arcsec, except for HC$_3$N.
The emission size of 35\arcsec was applied for HC$_3$N and its isotopologues and isomers (HCCNC and HNCCC) based on figure \ref{fig:map}.

The CS line was analyzed with two-velocity component fitting and each column density is summarized in table \ref{tab:column}.
We fitted the optically-thin hyperfine satellite lines of HC$_3$N ($J= 4-3$, $F= 4-4$ and $F= 3-3$) because the main satellite line ($J= 4- 3$, $F= 3-2$) is optically thick.
In the case of CH$_3$OH, we analyzed lines with $A$-type ($E_{\rm {up}}/k\approx2$ K) and $E$-type ($E_{\rm {up}}/k\approx15-29$ K) separately.
We summarized both results in table \ref{tab:column} with the notation of ($A$) and ($E$), respectively.

The SO line could not be fitted under the assumption of the LTE condition and the excitation temperature of 10 K because this line becomes optically thick ($\tau\approx1.5$). 
We thus analyzed this line with the non-LTE code RADEX \citep{2007A&A...468..627V}. 
The H$_2$ gas density of $9\times10^4$ cm$^{-3}$ was applied \citep{2013ApJ...778...34T}.
The background and gas kinetic temperatures were set at 2.73 K and 10 K, respectively.
The line width is 1.1 km\,s$^{-1}$, which was obtained from the Gaussian fitting (table \ref{tab:line}).
We derived errors by considering an absolute calibration error of 10\%.

Table \ref{tab:column} summarizes the derived column densities and fractional abundances with respect to H$_2$\footnote{Fractional Abundance $= N$($X$)/$N$(H$_2$) where $X$ denotes molecular species.}.
The column densities of H$_2$, $N$(H$_2$), in our target position was obtained from the fits file provided by the {\it {Herschel}} Gould Belt Survey \citep{2010A&A...518L.102A} Archive\footnote{\url{http://www.herschel.fr/cea/gouldbelt/en/Phocea/Vie_des_labos/Ast/ast_visu.php?id_ast=66}}.
The angular resolution of the $N$(H$_2$) fits file is $36.3\arcsec$.
The $N$(H$_2$) values at our target position are $8.1\times10^{22}$ cm$^{-2}$ and $7.8\times10^{22}$ cm$^{-2}$ with beam sizes of 30\arcsec and 35\arcsec, respectively.
The latter value is used only for HC$_3$N and its isotopologues and isomers (HCCNC and HNCCC).

\section{Discussions} \label{sec:dis}

\subsection{Comparisons of line width and velocity shifts} \label{sec:d1_3}

\begin{figure*}
 \begin{center}
  \includegraphics[bb = 0 20 940 610, width=\textwidth]{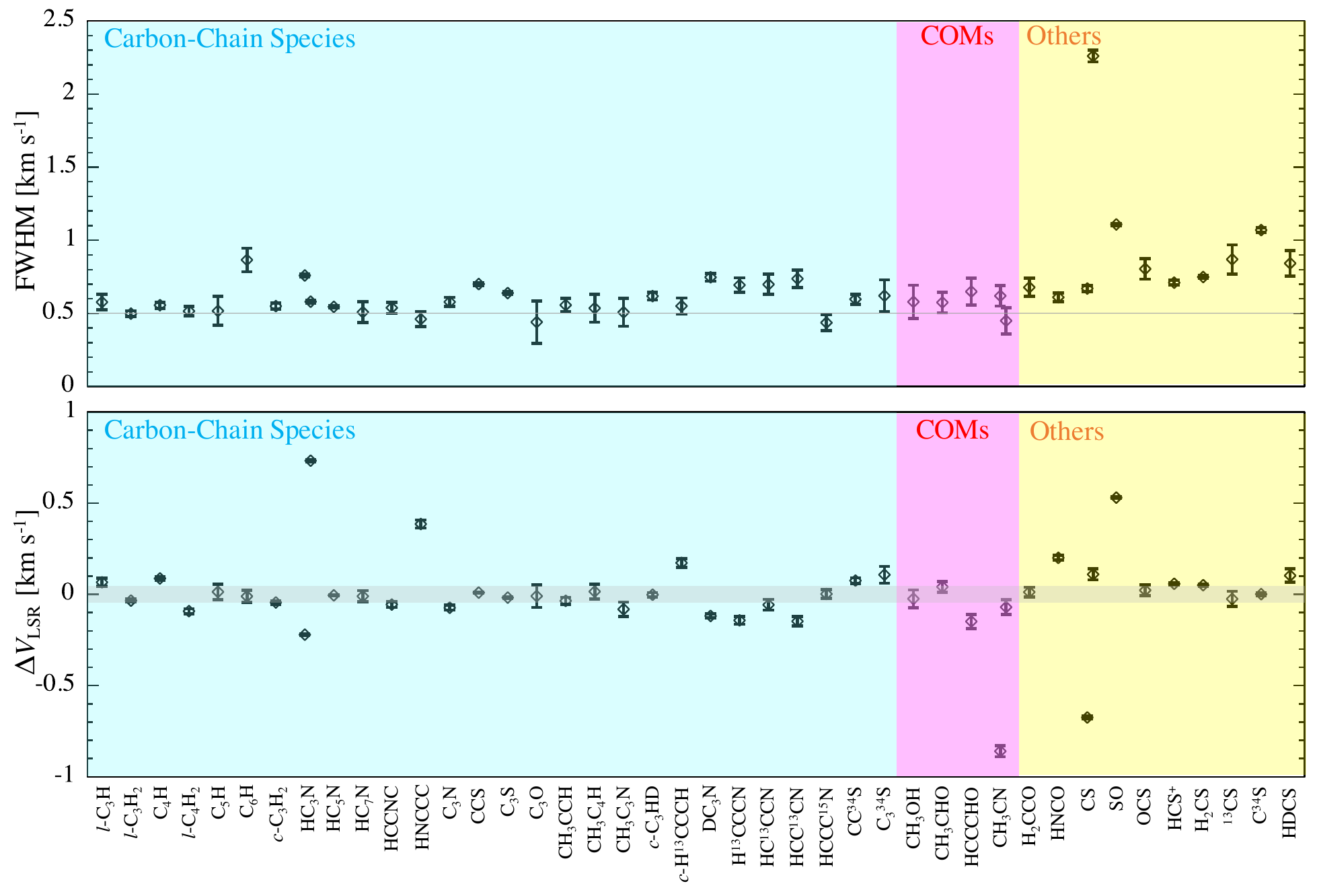} 
 \end{center}
\caption{Upper and lower panels show comparisons of line width (FWHM) and velocity shifts from the systemic velocity ($V_{\rm {LSR}}=7.5$ km\,s$^{-1}$), respectively. The highlighted region in the lower panel ($-0.05$ km\,s$^{-1}$ $\leq \Delta V_{\rm {LSR}} \leq$ $+0.05$ km\,s$^{-1}$) shows a velocity resolution of the spectra. {Alt text: Two panels. The vertical axes in the upper and lower panels indicate line width in the unit of kilometer per second the velocity shifts from the systemic velocity in the unit of kilometer per second, respectively. In the horizontal axis, the molecular species are listed.}}\label{fig:line_feature}
\end{figure*}

Figure \ref{fig:line_feature} shows comparisons of line width (FWHM) and velocity shifts from the systemic velocity ($\Delta V_{\rm {LSR}}$) derived by the Gaussian fitting.
In the case of HC$_3$N, CS, and CH$_3$CN, we plotted two velocity components.

As shown in the upper panel of figure \ref{fig:line_feature}, carbon-chain species show similar line widths around 0.5 km\,s$^{-1}$ -- 0.6 km\,s$^{-1}$.
These line widths are close to typical values in starless cores.
These narrow line widths suggest that these species trace cold gas.
Deuterium and $^{13}$C isotopologues of HC$_3$N show slightly wider line widths ($\sim0.7$ km\,s$^{-1}$), but they are comparable with one component of the HC$_3$N line ($\sim 0.76$ km\,s$^{-1}$).
As we discuss later, the HC$_3$N emission may come from outer turbulent regions.

Other species that are listed before CS show similar line widths with the carbon-chain species.
Hence, these species also trace the cold gas.
On the other hand, sulfur (S)-bearing species show wide line widths, especially CS and SO. 
One component of CS shows the largest value of 2.26 km\,s$^{-1}$.
These species may trace more turbulent regions. 
Or, the CS line may be associated with shock regions.

The lower panel shows the velocity shifts from the systemic velocity (7.5 km\,s$^{-1}$) determined by the C$^{18}$O line.
Almost all of the species show shifts less than $\pm 0.1$ km\,s$^{-1}$.
Three species that we applied two-velocity fitting (HC$_3$N, CH$_3$CN, and CS) show large shifts in one component, but the other components show small shifts less than 0.2 km\,s$^{-1}$.
In addition, the SO line shows a larger shift ($+ 0.5$ km\,s$^{-1}$).
The larger shifts in SO and CS lines also support that they trace turbulent regions.

To summarize, almost all of the detected species trace the cold gas which has the systemic velocity.
The exceptions are CS and SO which likely trace turbulent or shock regions.

\subsection{Comparisons of Carbon-Chain Abundances to TMC-1 CP} \label{sec:d1}

\begin{figure*}
 \begin{center}
  \includegraphics[bb = 0 17 620 745, scale = 0.55]{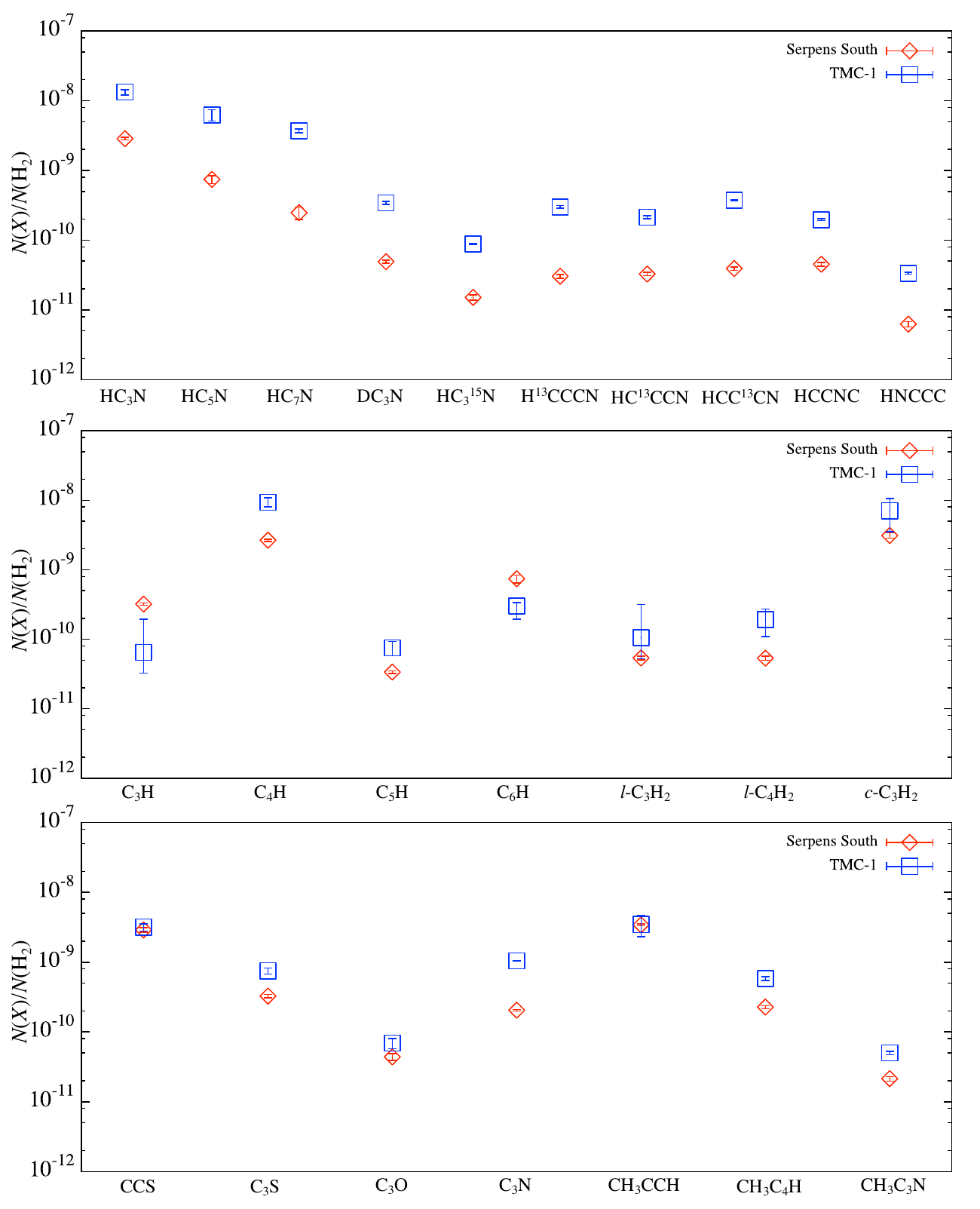} 
 \end{center}
\caption{Comparisons of abundances of carbon-chain species with respect to H$_2$ between Serpens South (red diamond) and TMC-1 CP (blue square). The top panel shows comparisons of cyanopolyynes, isotopologues, and isomers of HC$_3$N. The middle and bottom panels show comparisons of hydrocarbons (C$_n$H$_m$) and other carbon-chain species, respectively. Error bars indicate the standard deviation. The column densities at TMC-1 CP are taken from previous studies; \citet{2024A&A...682L..12T} for isotopologues of HC$_3$N; \citet{2024A&A...682L..13C} for HCCNC and HNCCC; \citet{2021A&A...650L...9C} for $l$-C$_4$H$_2$; \citet{2020A&A...641L...9C} for C$_3$N; \citet{1981ApJ...248L.113I} for CH$_3$CCH; \citet{2024Ap&SS.369...34T} for the other species. {Alt text: Three graphs. The abundance has no unit. The molecular species are listed in the vertical axis.}}\label{fig:com_carbon}
\end{figure*}

\begin{table}[h]
  \tbl{Carbon-chain abundance ratios of [TMC-1 CP]/[Serpens South]}{%
  \begin{tabular*}{35mm}{lc}
\hline 
Species & Ratio \\
\hline
HC$_3$N & 4.7	\\
HC$_5$N & 8.4	\\
HC$_7$N & 15.0	\\
DC$_3$N &	6.9	\\
HC$_3$$^{15}$N & 5.9 \\
H$^{13}$CCCN & 9.9 \\
HC$^{13}$CCN & 6.6 \\
HCC$^{13}$CN &9.6 \\
HCCNC & 4.4	\\
HNCCC &5.4 \\
$l$-C$_3$H & 0.2 \\
C$_4$H & 3.5 \\
C$_5$H & 2.2 \\
C$_6$H & 0.4 \\
$l$-C$_3$H$_2$ & 2.0 \\
$l$-C$_4$H$_2$ & 3.6 \\
$c$-C$_3$H$_2$ &2.3	\\
CCS & 1.1	\\
C$_3$S & 2.3 \\
C$_3$O & 1.6 \\
C$_3$N & 5.1 \\
CH$_3$CCH & 1.0	\\
CH$_3$C$_4$H & 2.6 \\
CH$_3$C$_3$N & 2.3 \\
\hline
\end{tabular*}}\label{tab:carbon_ratio}
\begin{tabnote}
\end{tabnote}
\end{table}

We compare fractional abundances of carbon-chain species with respect to H$_2$ in our target position to those at TMC-1 CP.
The $N$(H$_2$) value at TMC-1 was obtained from the fits file taken from the {\it {Herschel}} Gould Belt Survey Archive, and the value is $1.7\times10^{22}$ cm$^{-2}$. 
Table \ref{tab:carbon_ratio} summarizes the abundance ratios between TMC-1 CP and our target position in Serpens South. 

The top panel of figure \ref{fig:com_carbon} shows comparisons of cyanopolyynes (HC$_{2n+1}$N), and isotopologues and isomers of HC$_3$N.
As the chain becomes longer, the abundance of cyanopolyynes becomes lower in our target position compared to TMC-1 CP; the abundance ratios between TMC-1 CP and our target position are 4.7, 8.4, and 15.0 for HC$_3$N, HC$_5$N, and HC$_7$N, respectively (table \ref{tab:carbon_ratio}).
The abundances of DC$_3$N, HC$_3$$^{15}$N, HCCNC, and HNCCC in our target position are deficient by a factor of $\sim 4.4 - 10$.
The isotopic ratios and ratios among isomers will be discussed in sections \ref{sec:d2} and \ref{sec:d4}, respectively.
We will compare the abundances among the three $^{13}$C isotopologues of HC$_3$N to constrain its main formation pathway (see section \ref{sec:d3}).

The middle panel of figure \ref{fig:com_carbon} shows comparisons among hydrocarbons (C$_n$H$_m$) including $linear$ ($l$) and $cyclic$ ($c$) forms.
The abundance ratios between TMC-1 CP and our target position are between 0.2 -- 3.6.
Unlike cyanopolyynes, longer species do not clearly deplete compared to shorter carbon-chain species.
In the case of C$_3$H and C$_6$H, their abundances in our target position are higher than those in TMC-1 CP, which is an opposite trend for the other species.
However, this trend is not clear because they agree within their $2 \sigma$ errors.
The ratio of isomers ($l$- and $c$-C$_3$H$_2$) will be discussed in section \ref{sec:d4}.

The bottom panel of figure \ref{fig:com_carbon} shows comparisons of other carbon-chain species including sulfur (S), oxygen (O), and nitrogen (N) atoms, and terminated with -CH$_3$.
The differences between our target region and TMC-1 CP are a factor of 1.0 -- 5.1 (table \ref{tab:carbon_ratio}).
The abundances of CCS and CH$_3$CCH are consistent at the two cores.
The difference in the abundance of C$_3$N between our target core and TMC-1 CP is a factor of 5.1, which is comparable with HC$_3$N (4.7).
The differences in abundance between the two cores are generally larger for N-bearing carbon-chain molecules (cyanopolyynes and C$_3$N) compared to other species.

In summary, our target position shows the carbon-chain-rich feature.
Their abundances are consistent with the most carbon-chain-rich starless core TMC-1 CP within one order of magnitude, except for HC$_7$N.
These results mean that our target position contains the chemically young gas.

\subsection{Comparisons of Complex Organic Molecules among Starless Cores} \label{sec:d1_2}

Complex organic molecules (COMs) which consist of more than six atoms \citep{2009ARA&A..47..427H} have been found not only in hot cores/hot corinos around protostars but also starless cores ($e.g.,$ \cite{2016ApJ...830L...6J, 2023MNRAS.519.1601M,2024arXiv240811613S}).
In particular, CH$_3$OH and CH$_3$CHO are found to be prevalent in starless cores \citep{2020ApJ...891...73S}.
In this subsection, we compare the chemical compositions of COMs in our target position to other starless cores in low-mass star-forming regions.

\citet{2020ApJ...891...73S} conducted survey observations of CH$_3$OH and CH$_3$CHO toward 31 starless cores in the L1495-B218 filaments in the Taurus low-mass star-forming regions.
They derived the CH$_3$CHO/CH$_3$OH abundance ratios at 0.02 -- 0.26 in their target starless cores.
The ratios at our target position are calculated as $0.042\pm0.003$ and $0.090\pm0.007$, using the $A$-type and $E$-type transitions of CH$_3$OH, respectively.
Thus, the ratios at our target position are within the range derived in \citet{2020ApJ...891...73S}.
No clear difference between our target position and the Taurus region may suggest that W40 and the nearby cluster do not strongly affect the abundances of CH$_3$OH and CH$_3$CHO.

\begin{figure*}
 \begin{center}
  \includegraphics[bb = 0 17 520 240, scale = 0.8]{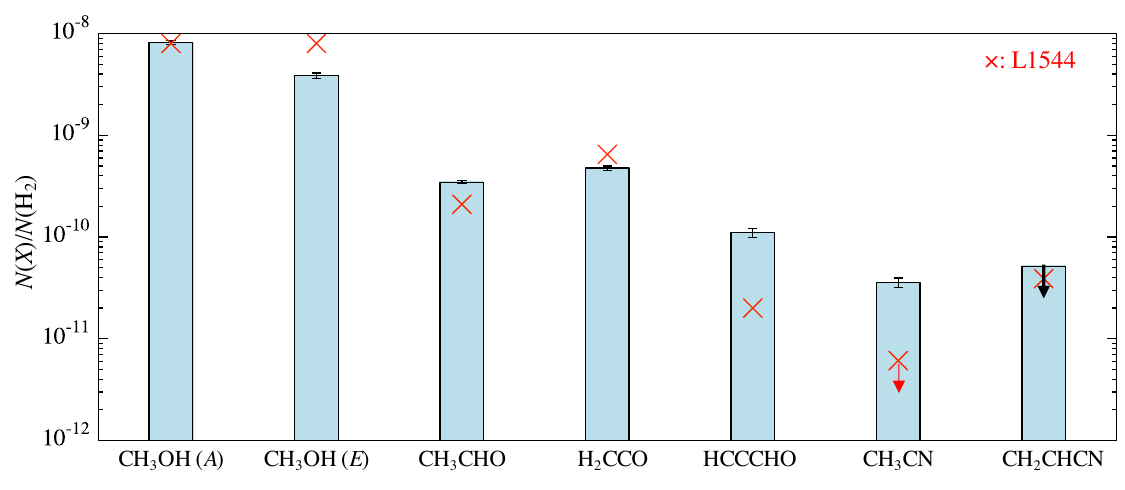} 
 \end{center}
\caption{Comparisons of abundances of COMs and H$_2$CCO between our target position (light-blue bar) and the CH$_3$OH peak in L1544 (red cross) \citep{2016ApJ...830L...6J, 2021ApJ...917...44J}. {Alt text: One graph. The vertical axis has no unit. The molecular species are listed in the horizontal axis.}}\label{fig:com_COM}
\end{figure*}

\citet{2023MNRAS.519.1601M} conducted observations toward the young starless core L1517B and compared the results to other more evolved starless cores, L1498 and L1544.
They proposed that the chemical complexity increases along the evolution of cores, and N-bearing COMs form first followed by the formation of the O-bearing COMs. 
\citet{2023MNRAS.519.1601M} did not detect CH$_2$CHCN and CH$_3$CHO, both of which are tentatively detected or detected in our target position.
CH$_2$CHCN has been detected in L1544 and L1498, while CH$_3$CHO has been detected in L1544.
Hence, our target position likely has similar chemical characteristics with L1544, which is considered to be the most evolved prestellar core with a kernel structure \citep{2022ApJ...929...13C}.
Figure \ref{fig:com_COM} shows comparisons of COMs in our target position and L1544.
The derived abundances in our target position are close to those in L1544, except for HCCCHO and CH$_3$CN.

The abundances of CH$_3$OH and CH$_3$CHO with respect to H$_2$ are approximately $3.0\times10^{-9}$ and $1.4\times10^{-10}$, respectively, in TMC-1 \citep{2018ApJ...854..116S}\footnote{These abundances are sums of blue and red components.}.
Their abundances in our target positions are higher than those in TMC-1 CP by a factor of $\sim1.3-2.7$, whereas the differences between our target position and L1544 are $\sim0.5-1.6$.
Therefore, the abundances of these two oxygen-bearing COMs in our target position are closer to those in L1544, rather than TMC-1 CP.

In summary, we could not find significant differences in the chemical composition of COMs between our target position in Serpens South and quiescent low-mass star-forming regions.
The abundances of COMs at our target position are significantly similar to those of L1544, which is the evolved prestellar core.


\subsection{Isotopic ratios} \label{sec:d2}

\begin{table}[h]
  \tbl{Isotopic ratio at the starless core in Serpens South}{%
  \begin{tabular}{lcc}
\hline 
Species & Ratio & Error\footnotemark[$*$]  \\
\hline
\multicolumn{3}{l}{\bf {D/H ratios}}  \\
DC$_3$N/HC$_3$N & 1.72 \% & 0.12 \% \\
$c$-C$_3$HD/$c$-C$_3$H$_2$ & 7.8 \%	& 0.6 \%  \\
HDCS/H$_2$CS &	11.3 \%	& 0.5 \% \\
\hline
\multicolumn{3}{l}{\bf {$^{12}$C/$^{13}$C ratios}} \\			
HC$_3$N/H$^{13}$CCCN & 94 & 7 \\
HC$_3$N/HC$^{13}$CCN & 87 & 6 \\
HC$_3$N/HCC$^{13}$CN & 73 & 5  \\
$c$-C$_3$H$_2$/$c$-H$^{13}$CCCH\footnotemark[$\dag$] & 30 & 4 \\
HC$_5$N/H$^{13}$CCCCCN	& $>50$ &  \\
HC$_5$N/HC$^{13}$CCCCN	& $>77$ &  \\
HC$_5$N/HCC$^{13}$CCCN	&$>64$ &  \\
HC$_5$N/HCCC$^{13}$CCN	& $>51$	&  \\
HC$_5$N/HCCCC$^{13}$CN	& $>76$	&  \\
\hline
\multicolumn{3}{l}{\bf {$^{14}$N/$^{15}$N ratios}} \\
HC$_3$N/HC$_3$$^{15}$N & 189	& 19 \\
\hline
\multicolumn{3}{l}{\bf {$^{32}$S/$^{34}$S ratios}}  \\			
CCS/CC$^{34}$S	& 36 & 4   \\
C$_3$S/C$_3$$^{34}$S & 23 & 3  \\
H$_2$CS/H$_2$C$^{34}$S & $>21$ &  \\
\hline
\end{tabular}}\label{tab:isotope_ratio}
\begin{tabnote}
\footnotemark[$*$] Errors indicate the standard deviation.  \\ 
\footnotemark[$\dag$] This is the off-axis $^{13}$C isotopologue and the expected ratio is 30 -- 35 if the local elemental $^{12}$C/$^{13}$C ratio is 60 -- 70 \citep{2005ApJ...634.1126M}.
\end{tabnote}
\end{table}

We have detected some isotopomers of D, $^{13}$C, $^{15}$N, and $^{34}$S.
Table \ref{tab:isotope_ratio} summarizes the isotope ratios of each species at our target position. 
We indicate the ratios as the lower limits if the isotope species are tentatively detected (S/N$\approx3$).

\subsubsection{Deuterium Fractionation} \label{sec:Deuterium}

Molecular species usually show much higher D/H ratios compared to the elemental D/H ratio ($\sim1.5\times10^{-5}$; \cite{2003SSRv..106...49L}).
Deuterium fractionation is one of the most important tracers to investigate chemical inheritance.
We detect three deuterium species in our target position; DC$_3$N, $c$-C$_3$HD, and HDCS.
The D/H ratios of these species are $1.72\pm0.12$ \%, $7.8\pm0.6$ \%, and $11.3\pm0.5$ \%, respectively.
The D/H ratio of H$_2$CS is higher than the carbon-chain species.
This is related to their formation mechanisms; the two carbon-chain species form only in the gas phase, while H$_2$CS could be produced both in the gas phase and on dust grains \citep{2022A&A...661A.111S}.

The HDCS/H$_2$CS at TMC-1 CP was derived to be $2.0 \pm 0.5$ \% \citep{1997ApJ...491L..63M}.
Hence, the D/H ratio of H$_2$CS at our target position is significantly higher than that at TMC-1 CP.
\citet{2022A&A...661A.111S} derived the HDCS/H$_2$CS ratio at the dust peak in the prestellar core L1544 to be $12\pm2$ \%, which is comparable with our target position. 
They also showed results of the gas-grain chemical simulations for the HDCS/H$_2$CS 
ratios in starless cores.
The high HDCS/H$_2$CS ratio in our target position ($11.3 \pm 0.5$ \%) is 
reproduced in their gas-grain chemical simulations around $10^5$ yr (a model for $n$(H$_2$)$=10^5$ cm$^{-3}$, figure 3 in \cite{2022A&A...661A.111S}).
The similar HDCS/H$_2$CS ratios imply that our target position chemically evolves to a similar degree with L1544.
This is consistent with the conclusion from the comparisons of COMs (section \ref{sec:d1_2}).

The $c$-C$_3$HD/$c$-C$_3$H$_2$ ratios at TMC-1 CP and the continuum peak in L1544 were derived to be 4.75 \% \citep{2001ApJS..136..579T} and 17 \% \citep{2022A&A...661A.111S}, respectively.
The D/H ratio of $c$-C$_3$H$_2$ in our target position is between TMC-1 CP and L1544. 
This seems to suggest that the $c$-C$_3$H$_2$ emission in our target position comes from relatively chemically young gas.
The DC$_3$N/HC$_3$N ratio at L1544 was derived to be 1.5 \% -- 9.3 \%, depending on assumed excitation temperatures (5 K and 10 K; \cite{2020A&A...633A.118L}), and the ratio at TMC-1 CP was derived to be $1.74 \pm 0.03$ \% \citep{2024A&A...682L..12T}.
The ratio at our target position ($1.72 \pm 0.12$ \%) is close to the lower limit of L1544 and comparable with TMC-1 CP.
Again, this implies that the HC$_3$N emission comes from the chemically young gas.
The different D/H ratios between $c$-C$_3$H$_2$ and HC$_3$N at our target position suggest that these two carbon-chain species trace slightly different regions; HC$_3$N traces chemically younger gas.

Next, we compare the D/H ratios of the carbon-chain species in other cores.
\citet{2018ApJ...863..126C} observed $c$-C$_3$H$_2$ and its deuterium 
isotopologue toward 11 starless or prestellar cores.
They derived the D/H ratios to be 8 \% -- 13 \%, which are consistent with our result.
\citet{2020MNRAS.496.1990R} conducted survey observations and derive the DC$_3$N/HC$_3$N ratios in high-mass star-forming regions.
They derived the DC$_3$N/HC$_3$N ratios in high-mass cores to be 0.3 \% -- 2.2 \%, which is lower than prestellar cores in low-mass dark clouds (3 \% -- 13 \%, see figure 7 in \cite{2020MNRAS.496.1990R}).
Since they observed using the IRAM 30 m telescope, most of the emission comes from outer envelopes rather than protostellar cores.
\citet{2020MNRAS.496.1990R} concluded that the deuterium fractionation of HC$_3$N gives an estimation of the deuteration factor prior to the formation of dense gas by comparing the deuterium fractionation of the other species.
The DC$_3$N/HC$_3$N ratio in our target region (1.72 \%) is close to those in high-mass cores (see figure 7 in \cite{2020MNRAS.496.1990R}). 
This implies that the HC$_3$N emission in our target position comes from outer envelope gas which does not chemically evolve so much.

The different D/H ratios in HC$_3$N, $c$-C$_3$H$_2$, and H$_2$CS imply that these three species trace different regions.
The densest and coldest region is traced by H$_2$CS, whereas $c$-C$_3$H$_2$ and HC$_3$N trace the outer diffuse gas that surrounds the densest core.
The HC$_3$N emission likely traces outer regions compared to $c$-C$_3$H$_2$.
In fact, HC$_3$N shows two velocity components and a wider line width (see section \ref{sec:d1_3}), which indicates that HC$_3$N traces more turbulent regions.
This is also consistent with the scenario proposed by \citet{2020MNRAS.496.1990R}; deuterium fractionation of HC$_3$N gives an estimation of the deuteration factor prior to the formation of dense gas.

\subsubsection{Carbon Isotopic Fractionation} \label{sec:13Carbon}

The elemental $^{12}$C/$^{13}$C ratio in the local interstellar medium (ISM) is around 60 -- 70 \citep{2005ApJ...634.1126M}.
However, carbon-chain species show higher $^{12}$C/$^{13}$C ratios in cold starless cores ($e.g.,$ \cite{1998A&A...329.1156T,2016ApJ...817..147T,2019ApJ...884..167T}).
This is known as the dilution of the $^{13}$C species, which is caused by the lock of $^{13}$C in CO molecules (for a review see \cite{2024Ap&SS.369...34T}).
The degree of dilution likely differs among the species; the $^{12}$C/$^{13}$C ratios in cyanopolyynes are lower than CCH.

If heating from the W40 H$_{\rm {II}}$ region or the protostars in the central cluster of Serpens South affects the chemistry at our target position, it is expected that we cannot see the dilution of the $^{13}$C species in carbon-chain molecules due to the following two reasons; (1) $^{13}$C is not locked in CO in warm regions ($i.e.,$ $T \approx 35$ K), and (2) the UV radiation from W40 could destroy $^{13}$CO more efficiently compared to $^{12}$CO, namely the isotope-selective photodissociation.
In these conditions, the gas-phase abundances of $^{13}$C and $^{13}$C$^+$ could increase and they are available for the formation of carbon-chain species.

In our target position, the $^{12}$C/$^{13}$C ratios of HC$_3$N are approximately 70 -- 100, which is almost comparable to or slightly higher than the elemental ratio in the local ISM.
In TMC-1 CP, the $^{12}$C/$^{13}$C ratios were derived to be $79 \pm 11$, $75 \pm 10$, and $55 \pm 7$ for H$^{13}$CCCN, HC$^{13}$CCN, and HCC$^{13}$CN, respectively \citep{1998A&A...329.1156T}.
The ratios in our target region are consistent with or higher than TMC-1 CP.

In the case of $c$-C$_3$H$_2$, we have detected the off-axis $^{13}$C isotopologues, which have two symmetry structures. 
Therefore, the expected $c$-C$_3$H$_2$/$c$-H$^{13}$CCCH ratio is 30 -- 35, if the elemental $^{12}$C/$^{13}$C ratio is 60 -- 70.
The $c$-C$_3$H$_2$/$c$-H$^{13}$CCCH ratio at our target position is derived to be $30 \pm 4$, which is well consistent with the predicted value.

We tentatively detected the five $^{13}$C isotopologues, and we could not constrain its $^{12}$C/$^{13}$C ratios.
The derived lower limits cannot suggest the dilution of the $^{13}$C species, and we need high-sensitivity observations.

In summary, we could not see clear evidence for the $^{13}$C dilution of the carbon-chain species in our target position.
We need to observe species that usually show heavy dilution ($e.g.,$ CCH) in our target position to assess the effects of the heating and the UV radiation on the carbon isotopic fractionation.

\subsubsection{Nitrogen Isotopic Fractionation} \label{sec:15Nitrogen}

The $^{14}$N/$^{15}$N ratio in the solar wind is 440 \citep{2011Sci...332.1533M} and that in the atmosphere of Jupiter is 450 \citep{2004Icar..172...50F}.
On the other hand, the terrestrial atmosphere (272; \cite{1950PhRv...77..789N}) and comets in the solar system also show lower values ($e.g.,$ \cite{2004ApJ...613L.161J}).
The nitrogen fractionation has been focused on because it can trace the evolution of primordial Solar System material \citep{2014A&A...572A..24W, 2018MNRAS.474.3720W} as well as the deuterium fractionation.

In our observations, we detected only one nitrogen isotopologue, HC$_3$$^{15}$N, and the HC$_3$N/HC$_3$$^{15}$N ratio is derived to be $189 \pm 19$.
\citet{2017PASJ...69L...7T} derived the ratio to be $257 \pm 54$ at TMC-1 CP.
This result has been confirmed in a recent study ($235 \pm 46$; \cite{2024A&A...682L..12T}).
\citet{2018MNRAS.480.1174H} derived the HC$_3$N/HC$_3$$^{15}$N ratio at $400 \pm 20$ in L1544. 
The HC$_3$N/HC$_3$$^{15}$N ratio at our target position is lower than those in TMC-1 CP and L1544.

As we discuss in section \ref{sec:d3}, CN is considered to be a precursor of HC$_3$N and its nitrogen isotopic fractionation could be inherited to HC$_3$N.
The low $^{14}$N/$^{15}$N ratio in CN could be caused by the isotope-selective photodissociation of the N$_2$ molecules \citep{2018ApJ...857..105F,2022A&A...664L...2S}.
\citet{2018ApJ...857..105F} found the concentration of $^{15}$N in CN around the transition phase from the atomic nitrogen to N$_2$ ($A_v\approx1.8$ mag).
The C$^{14}$N/C$^{15}$N ratios could become lower than 200 with visual extinction of $\sim 1.4$ mag -- 2.3 mag.
The visual extinction at the observed position is calculated as $\sim 75$ mag using the relationship with $N_{\rm {H}}$ derived by \citet{2017MNRAS.471.3494Z} ($A_v = N_{\rm {H}}/[(2.08\pm0.02)\times10^{21}$] mag).
However, as discussed in section \ref{sec:Deuterium}, the emission of HC$_3$N likely comes from outer regions with a lower visual extinction. 
The low-$A_v$ condition could also enhance the CN abundance owing to the photodissociation of HCN, which can lead efficient formation of HC$_3$N from CN (see section \ref{sec:d3}).
\citet{2022A&A...664L...2S} found that the $^{14}$N/$^{15}$N ratios of CN and HCN decrease toward the southern edge of L1544 where the interstellar radiation field irradiates, and they proposed that isotope-selective photodissociation has a strong effect on the nitrogen fractionation across the prestellar cores.
Our result of the low $^{14}$N/$^{15}$N ratio suggests that the UV radiation from the W40 H$_{\rm {II}}$ region may affect the nitrogen fractionation.
To assess this point further, we need mapping observations to obtain spatial distributions of the $^{14}$N/$^{15}$N ratios.

\subsubsection{Sulfur Isotopic Fractionation} \label{sec:34Sulfur}

The $^{32}$S/$^{34}$S elemental isotope ratio in the local ISM was derived to be $24 \pm 4$ \citep{2023A&A...670A..98Y}. 
We detected sulfur isotopologues of two carbon-chain species, CCS and C$_3$S.
Their $^{32}$S/$^{34}$S ratios are calculated as $36 \pm 4$ and $23 \pm 3$, respectively.
The $^{32}$S/$^{34}$S ratios of the carbon-chain species in our target region agree with the local ISM value within their $3 \sigma$ errors. 
These ratios at TMC-1 CP are calculated as $11 \pm 2$ and $25 \pm 3$, respectively \citep{2021A&A...646L...3C}.
Hence, the $^{32}$S/$^{34}$S ratio of CCS at TMC-1 CP is lower but that of C$_3$S is consistent with our result.
We tentatively detected H$_2$C$^{34}$S. 
The lower limit of the $^{32}$S/$^{34}$S ratio of H$_2$CS is 21, and we could not find the isotope anomaly in H$_2$CS.

Two isotopologues of CS, $^{13}$CS and C$^{34}$S, are detected in our observations. 
The observed C$^{34}$S/$^{13}$CS ratio is $1.5 \pm 0.2$.
The C$^{34}$S/$^{13}$CS ratio in the local ISM is calculated as $2.8 \pm 0.6$, which is derived by the following formulae adopting the results of the CS isotopologues \citep{2023A&A...670A..98Y}; 
\begin{equation} \label{equ:isoratio}
\frac{\mathrm{C}^{34}\mathrm{S}}{^{13}\mathrm{CS}} = \frac{^{12}\mathrm{C}}{^{13}\mathrm{C}} \times \frac{^{34}\mathrm{S}}{^{32}\mathrm{S}}.
\end{equation}
The observed C$^{34}$S/$^{13}$CS ratio is slightly lower than the predicted value, but it is still consistent within their errors.
The C$^{34}$S/$^{13}$CS ratios in envelopes around intermediate-mass protostars show almost constant values and agree with the local ISM value (Taniguchi et al., {\it {submitted to \aap}}), which is consistent with our results.
This isotope ratio may not be affected by physical conditions ($e.g.,$ temperature, density) significantly.

The observed sulfur isotopic ratios agree with the predicted values within the $3 \sigma$ errors.
This means that $^{34}$S is not heavily diluted or concentrated in the detected species.

\subsection{$^{13}$C Isotopic Fractionation of HC$_3$N} \label{sec:d3}

Main formation pathways of carbon-chain species can be constrained by their $^{13}$C isotopic fractionation (for a review see \cite{2024Ap&SS.369...34T}).
Main formation pathways of HC$_3$N have been investigated in several starless cores \citep{1998A&A...329.1156T,2017ApJ...846...46T} and protostellar cores \citep{2016ApJ...833..291A, 2016ApJ...830..106T, 2021ApJ...908..100T}.
Until now, three patterns have been discovered; abundance ratios of [H$^{13}$CCCN] : [HC$^{13}$CCN] : [HCC$^{13}$CN] are (1) $1:1:x$, (2) $1:y:z$, and (3) $\approx 1:1:1$, where $x$, $y$, and $z$ are arbitrary values.
These fractionation patterns imply that the main formation pathways of HC$_3$N are (1) C$_2$H$_2$ + CN, (2) CCH + HNC, and (3) HC$_3$NH$^+$ + e$^-$, respectively.
In this subsection, we investigate $^{13}$C isotopic fractionation of HC$_3$N and constrain its main formation pathway.

Using the column densities of the three $^{13}$C isotopomers, we obtained the abundance ratios at our target position as follows: [H$^{13}$CCCN] : [HC$^{13}$CCN] : [HCC$^{13}$CN] $= 1.00:1.07 (0.09):1.29 (0.11)$, where numbers in parentheses are the standard deviation.
The results mean that the main formation pathway of HC$_3$N is the reaction between C$_2$H$_2$ and CN.
The result in our target position is consistent with those in TMC-1 CP \citep{1998A&A...329.1156T} and L1521B \citep{2017ApJ...846...46T}.
We thus conclude that the formation process of HC$_3$N in our target position is the same as other early starless cores in the Taurus region (TMC-1CP and L1521B).
These results are consistent with our previous discussion that the HC$_3$N emission traces the chemically young gas.
The UV radiation from W40 may produce the high abundance of CN in the outer chemically young gas (section \ref{sec:15Nitrogen}), leading to a larger contribution of the reaction of ``C$_2$H$_2$ + CN''.

\subsection{Abundance ratios among isomers} \label{sec:d4}

Two species of HCCNC and HNCCC are isomers of HC$_3$N, and $c$-C$_3$H$_2$ is an isomer of $l$-C$_3$H$_2$.
In this subsection, we compare the abundance ratios among these isomers.

\subsubsection{HC$_3$N, HCCNC, and HNCCC}

Three isomers of HC$_3$N, HCCNC, and HNCCC were classically considered to be formed via the electron recombination reaction of HC$_3$NH$^+$, and then the branching ratios of the recombination reaction were a key for understanding their abundance ratios. 
\citet{2019A&A...625A..91V} showed that this recombination reaction has the same branching ratios for HC$_3$N and HNCCC (24\%), but the route for HCCNC is less efficient (4\%).
Instead, HCCNC could be formed via the electron recombination reaction of HCCNCH$^+$, which is formed by ``C$^+$ + CH$_3$CN'' \citep{2024A&A...682L..13C}.
Moreover, as we discuss in section \ref{sec:d3}, the main formation route of HC$_3$N is the reaction between C$_2$H$_2$ and CN.
Hence, these three isomers are likely formed by different pathways and their abundance ratios could be independent from the branching ratios of the electron recombination reaction of HC$_3$NH$^+$.

We calculated the column density ratios of HC$_3$N/HCCNC, HC$_3$N/HNCCC, and HCCNC/HNCCC.
They are determined to be $63\pm5$, $456\pm45$, and $7\pm1$, respectively.
These ratios at TMC-1 CP were reported as 56, 328, and 6, respectively \citep{2024A&A...682L..13C}.
Therefore, HNCCC is deficient with respect to HC$_3$N in our target position compared to TMC-1 CP, whereas HCCNC does not significantly decrease.

The lower abundance of HNCCC with respect to HC$_3$N implies a smaller contribution of the electron recombination reaction of HC$_3$NH$^+$ to the formation of HC$_3$N.
This is consistent with the results that we could not detect HC$_3$NH$^+$, which was detected in TMC-1 CP \citep{2004PASJ...56...69K}.
We checked the noise levels at the frequency where the rotational transition line of this ion lies (34.631914 GHz for the $J=4-3$ line).
The noise levels of our data are around 10 mK.
If the ion has a similar column density as TMC-1 CP, we expect to detect this ion with S/N$\approx 3.5-4$ \citep{2004PASJ...56...69K}.
Thus, HC$_3$NH$^+$ is deficient in our target position.
As we discuss in section \ref{sec:d3}, the main formation route of HC$_3$N is the reaction of ``C$_2$H$_2$ + CN''.
This is other evidence for a smaller contribution of the electron recombination reaction of HC$_3$NH$^+$.
The reaction of ``C$_2$H$_2$ + CN'' was also suggested to be the main formation route of HC$_3$N at TMC-1 CP \citep{1998A&A...329.1156T}, but this reaction likely has a larger contribution in our target position compared to TMC-1 CP.

\subsubsection{$c$-C$_3$H$_2$ and $l$-C$_3$H$_2$}

The $cyclic$-to-$linear$ ratios (hereafter $c/l$ ratio) of C$_3$H$_2$ isomers have been investigated by theoretical and observational studies ($e.g.,$ \cite{2016A&A...591L...1S}, \cite{2017MNRAS.470.4075L}).
The $c/l$ ratios in prestellar cores were derived to be $\sim30-110$ \citep{2016A&A...591L...1S, 2017MNRAS.470.4075L}.
\citet {2017MNRAS.470.4075L} derived the ratio in TMC-1 CP to be 67, whereas the ratio in L1544 shows a lower value of 32 \citep{2016A&A...586A.110S}.
The different ratios likely suggest different density regions where both species trace.
\citet{2017MNRAS.470.4075L} proposed that isomerization reactions of ``$l$-C$_3$H$_2$ + H $\rightarrow$ $c$-C$_3$H$_2$ + H'' and ``$t$-C$_3$H$_2$ + H $\rightarrow$ $c$-C$_3$H$_2$ + H'' are important for the high $c/l$ ratio in low-density conditions.

The observed $c/l$ ratio of C$_3$H$_2$ in our target position is $58 \pm 6$, which is consistent with the value at TMC-1 CP within $2 \sigma$ error.
The similar ratios in our target position and TMC-1 CP imply that these carbon-chain species in our target position trace a similar density region with TMC-1 CP, and a less dense region than L1544.

\subsection{Astrochemical characteristics of a {\it {pre-cluster clump}}}

Infrared dark clouds (IRDCs) are considered to be the birthplace of clusters.
However, most IRDCs already contain protostars, making it challenging to find clumps that are not currently undergoing star formation.
Our target position is located in the northern clump identified by \citet{2013ApJ...778...34T}.
This clump has a mass of 193 M$_{\Sol}$, which is comparable to the central Serpens South cluster-forming clump (232 M$_{\Sol}$).
According to the infrared observations, no protostars are identified in the northern clump \citep{2015A&A...584A..91K}.
Thus, this clump is likely to be a clump immediately before star-cluster formation.

Here, we consider the current state of the clump from an astrochemical point of view.
The key chemical characteristics we identified are as follows:
\begin{enumerate}
    \item The chemical compositions of carbon-chain species are similar to TMC-1 CP, suggesting that our target position possesses the chemically young gas (section \ref{sec:d1}).
    \item The chemical compositions of COMs are similar to L1544, which is the evolved prestellar core (section \ref{sec:d1_2}). The results mean that our target position is chemically evolved but before the onset of protostar formation.
    \item The high D/H ratio of H$_2$CS is derived at our target position ($11.3\pm0.5$ \%). This value is similar to that in L1544 (section \ref{sec:Deuterium}). This is additional evidence for the prestellar-core like chemical feature.
\end{enumerate}
The co-existence of chemically young and evolved components can be naturally explained if the higher density region (chemically more evolved) is surrounded by the lower density region (chemically less evolved). 
Since the clump is barely resolved by the beam of the 45 m telescope, the chemically evolved gas in the inner region and the outer chemically young gas can be observed simultaneously in its beam.
Based on these chemical characteristics, we conclude that the northern clump is a {\it {pre-cluster clump}} that has the potential to form a cluster but without any star formation activity at the moment.

Our findings propose one method to confirm whether it is really a {\it {pre-cluster clump}}.
If these two chemical characteristics, which are similar to starless and prestellar cores, co-exist in an IRDC clump without YSOs identified by the infrared regime, it would be a {\it {pre-cluster clump}}.
The chemical compositions are a strong tool for investigating the evolutionary stages of clumps.

\section{Summary} \label{sec:sum}

We conduct the Q-band (30 GHz -- 50 GHz) line survey observations toward a peak position of the emission of the carbon-chain species in the Serpens South cluster-forming region.
We use the eQ receiver installed in the Nobeyama 45 m radio telescope.
The main results of this paper are as follows.

\begin{enumerate}
\item Approximately 180 lines are detected in the observed frequency range. These lines are attributed to 52 species including isotopologues. We detect almost all of the species that \citet{2004PASJ...56...69K} detected in the same frequency range at TMC-1 CP, except for three species (CCO, HC$_3$NH$^+$, and CH$_2$CN).

\item The rotational temperatures of HC$_5$N, HC$_7$N, and C$_6$H are derived to be 8 K -- 10 K, which are slightly higher than a typical value in TMC-1 CP (6.5 K).

\item In the case of cyanopolyynes (HC$_{2n+1}$N), the carbon chain becomes longer, and their abundance becomes lower in our target position compared to TMC-1 CP.
However, we could not find such a trend in hydrocarbons (C$_n$H$_m$).

\item Several COMs are detected. Their abundance and chemical complexity in our target position are similar to those in the evolved prestellar core L1544. These results mean that our target position evolves at a similar stage to L1544. 

\item The D/H ratios of HC$_3$N, $c$-C$_3$H$_2$, and H$_2$CS are derived to be $1.72 \pm 0.12$\%, $7.8 \pm 0.6$\%, and $11.3 \pm 0.5$ \%, respectively. The HDCS/H$_2$CS ratio in our target position is similar to that in L1544, and both gas-phase and dust-grain formation processes contribute to the formation of H$_2$CS. On the other hand, the carbon-chain species, which are formed in the gas phase, show slightly lower D/H ratios. These results suggest that H$_2$CS (and HDCS) traces the coldest and densest core, whereas the carbon-chain species trace outer less dense gas.

\item We detect carbon ($^{13}$C), nitrogen ($^{15}$N), and sulfur ($^{34}$S) isotopologues. We do not recognize any significant isotope anomaly at our target position.

\item We obtain the $^{13}$C isotopic fractionation of HC$_3$N; [H$^{13}$CCCN] : [HC$^{13}$CCN] : [HCC$^{13}$CN] $= 1.00:1.07 (0.09):1.29 (0.11)$.
This fractionation pattern indicates that the main formation pathway of HC$_3$N is the reaction between C$_2$H$_2$ and CN.
This is the same result in the young starless cores TMC-1 CP and L1521B.

\item The observed $c$-C$_3$H$_2$/$l$-C$_3$H$_2$ ratio is derived to be $58 \pm 6$, which is similar to that in TMC-1 CP (67) but higher than L1544 (32). These suggest that the carbon-chain species in our target position exist in a similar density region with TMC-1 CP but are less dense compared to L1544.
\end{enumerate}

It is found that our target position is rich in carbon-chain species similar to TMC-1 CP when we focus on carbon-chain species.
On the other hand, the chemical complexity of COMs and the deuterium fractionation suggest similarities with the evolved prestellar core L1544.
These results mean that our target position contains the densest and coldest part whose chemistry is similar to L1544 surrounded by the chemically young gas that has similar chemical compositions to TMC-1 CP.
Based on these chemical features, the northern clump including our target position is found to be a {\it {pre-cluster clump}} without any current star formation activity.

\begin{ack}
We would like to express thankfulness to the staff of the Nobeyama Radio Observatory. 
The Nobeyama 45 m radio telescope is operated by Nobeyama Radio Observatory, a branch of National Astronomical Observatory of Japan.

K.T. is supported by JSPS KAKENHI grant Nos. JP20K14523, 21H01142, 24K17096, and 24H00252. 
K.T. is grateful for research funds for NAOJ Fellow to purchase the observing time of the Nobeyama 45 m radio telescope. 
F.N. is supported by JSPS KAKENHI grant No. JP23H01218.
T.S. is supported by JSPS KAKENHI grant No. JP22K02966.
\end{ack}

\appendix 
\section{Integrated-Intensity maps} \label{sec:a1}

\begin{figure*}
 \begin{center}
  \includegraphics[bb= 0 10 640 740, width=\textwidth]{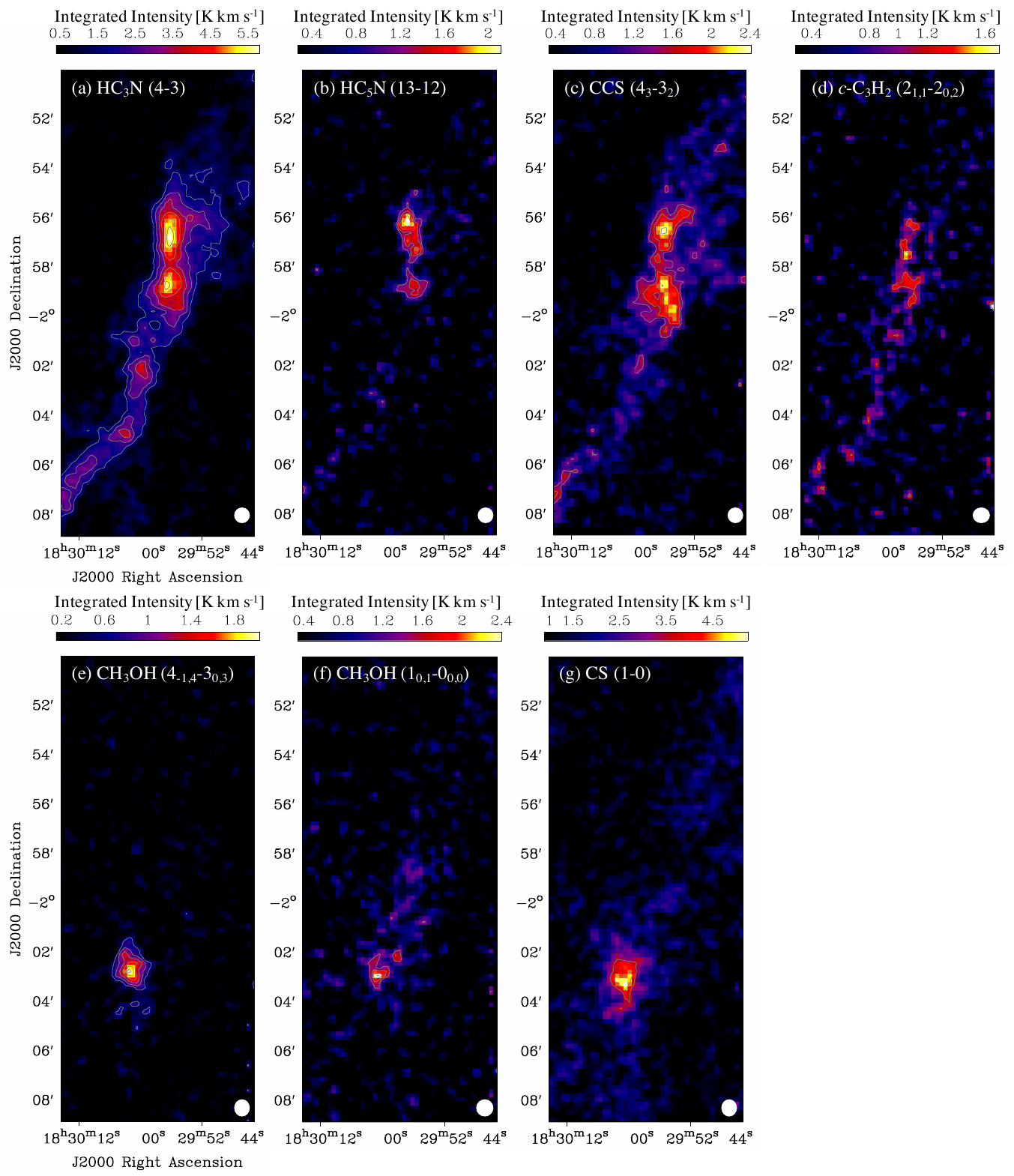} 
 \end{center}
\caption{Integrated-intensity maps of (a) HC$_3$N ($4-3$), (b) HC$_5$N ($13-12$), (c) CCS ($4_3-3_2$), (d) $c$-C$_3$H$_2$ ($2_{1,1}-2_{0,2}$), (e) CH$_3$OH ($4_{-1,4}-3_{0,3}$ $E$), (f) CH$_3$OH ($1_{0,1}-0_{0,0}$ $A$), and (g) CS ($1-0$). Noise levels of each panel are (a) 0.4 K\,km\,s$^{-1}$, (b) 0.27 K\,km\,s$^{-1}$, (c) 0.33 K\,km\,s$^{-1}$, (d) 0.29 K\,km\,s$^{-1}$, (e) 0.15 K\,km\,s$^{-1}$, (f) 0.34 K\,km\,s$^{-1}$, and (g) 0.89 K\,km\,s$^{-1}$, respectively. The contour levels are $4,6,8,10,12,13\sigma$ for panels (a) and (e), $4,5,6,7\sigma$ for panels (b), (c), and (f), and $4,5\sigma$ for panels (d) and (g). The white-filled circle at the bottom right shows a typical beam size of the Q-band (37\arcsec). {Alt text: Seven two-dimensional maps in the right ascension-declination coordinate. The unit is Kelvin kilometer per second.}}\label{fig:map_all}
\end{figure*}

Figure \ref{fig:map_all} shows integrated-intensity maps obtained with the eQ receiver.
Carbon-chain species are concentrated at the northern clumps, whereas CH$_3$OH and CS show strong emission peaks at Serpens South central clump harboring protostars \citep{2013ApJ...778...34T}.
The S/N ratio of $c$-C$_3$H$_2$ (panel d) is not high enough to obtain its spatial distribution.

\section{Close-up spectral figures} \label{sec:a2}

Figure \ref{fig:s1} shows close-up spectra.
In the 2SB mode observations, few tens of channels at the edge of each array in USB are missing due to the system of the data reduction software NEWSTAR\footnote{Provided by the Nobeyama Radio Observatory.}.
Then, some frequency bands show no data ($T_{\rm {A}}^*=0$ K).

Figure \ref{fig:3spec} show spectra of SO ($J_N=1_0-0_1$), HC$_3$N ($J=4-3$) and CS ($J=1-0$). The black lines indicate the observed spectra and red curves show the Gaussian fitting results.
The main component of the HC$_3$N line and CS line were fitted with two-velocity components, and the indicated curves are the synthesized ones.

\begin{figure*}
 \begin{center}
  \includegraphics[bb= 0 15 600 900, scale = 0.72]{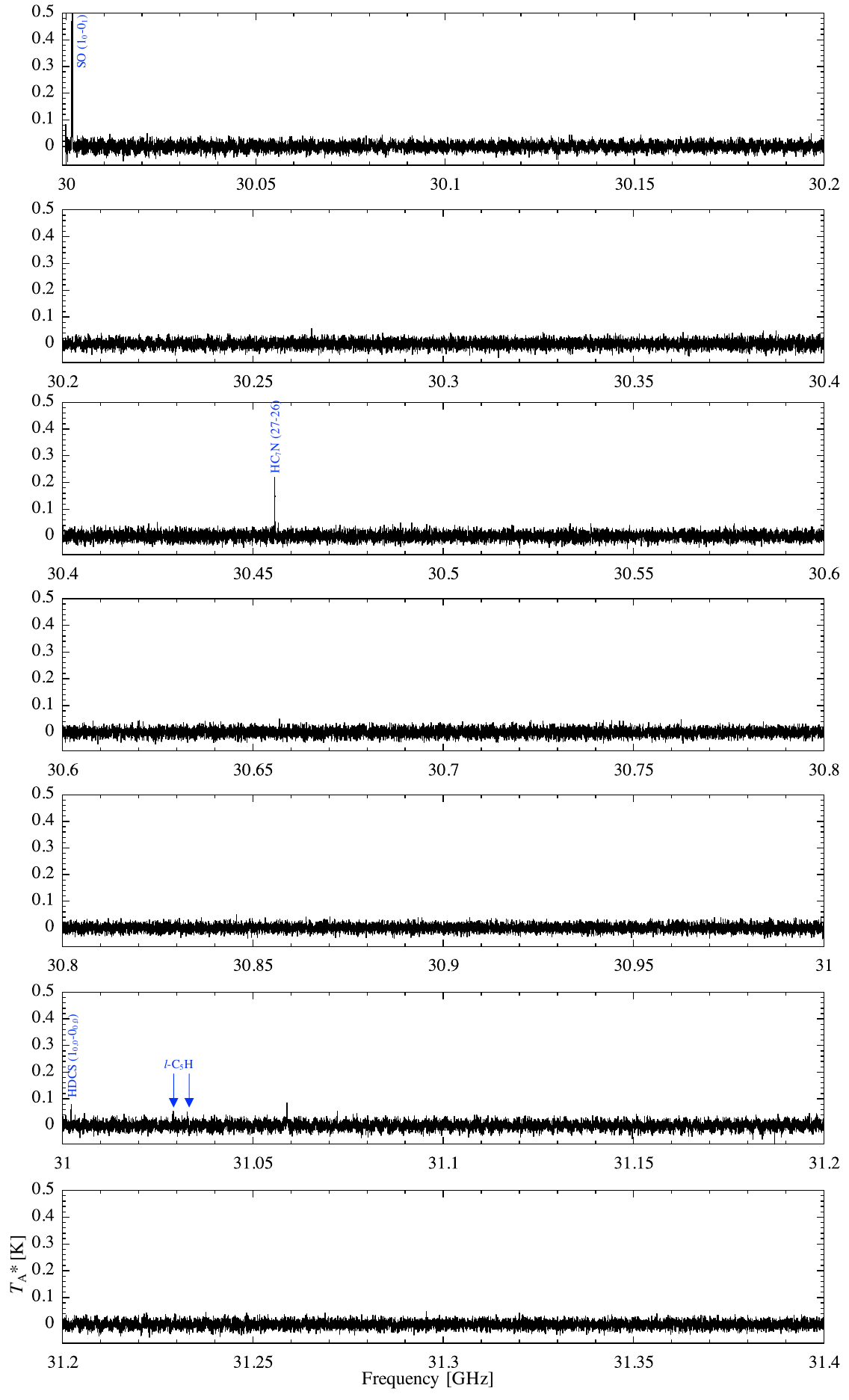} 
 \end{center}
\caption{Close-up spectra. {Alt text: Seven spectral plots. The vertical axis shows the antenna temperature in the unit of Kelvin, and the horizontal axis shows the frequency in the unit of gigahertz.}}\label{fig:s1}
\end{figure*}

\begin{figure*}
\setcounter{figure}{7}
 \begin{center}
  \includegraphics[bb= 0 15 600 900, scale = 0.72]{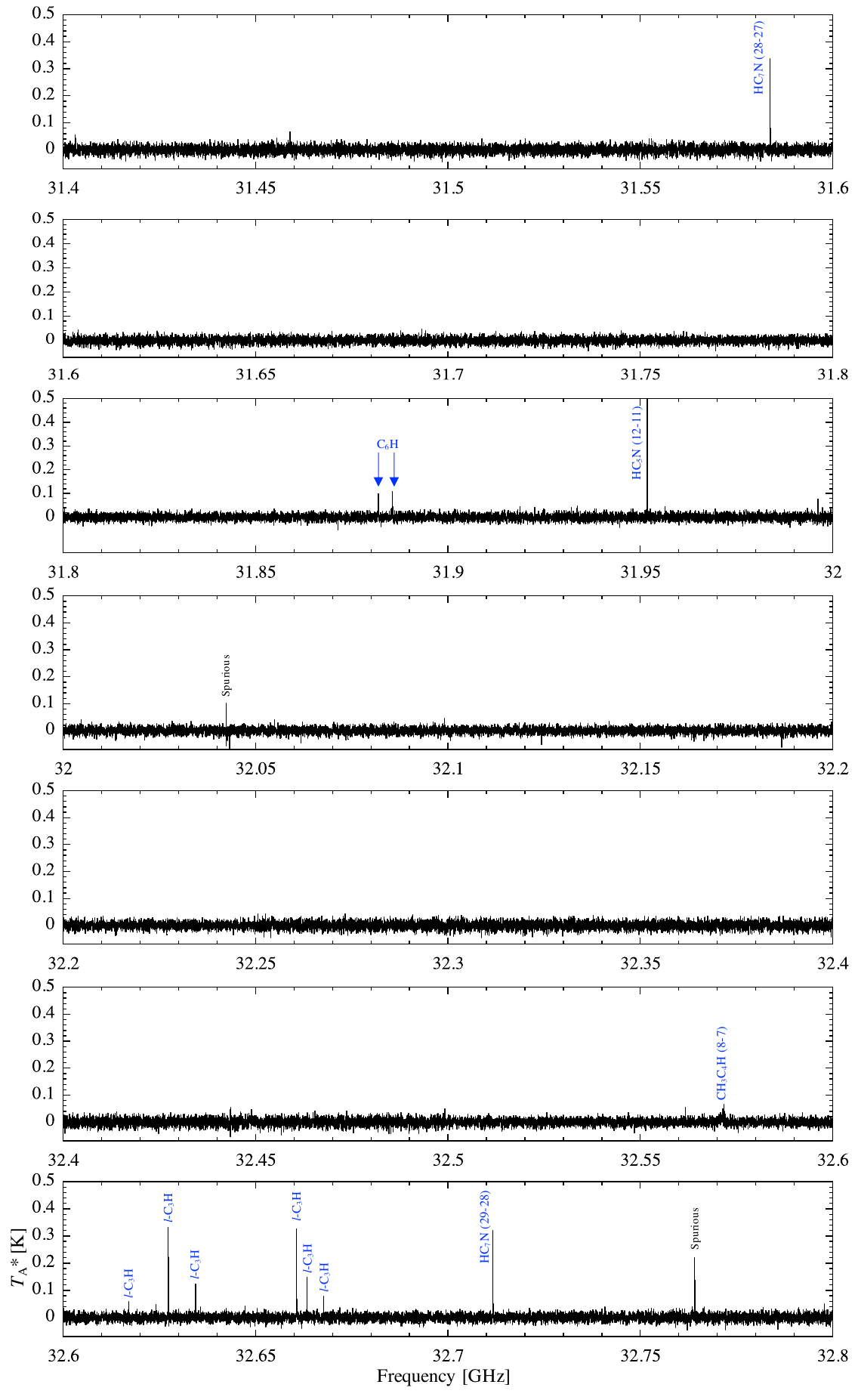} 
 \end{center}
\caption{Close-up spectra (Continued). {Alt text: Seven spectral plots. The vertical axis shows the antenna temperature in the unit of Kelvin, and the horizontal axis shows the frequency in the unit of gigahertz.}}\label{fig:s2}
\end{figure*}

\begin{figure*}
\setcounter{figure}{7}
 \begin{center}
  \includegraphics[bb= 0 15 600 900, scale = 0.72]{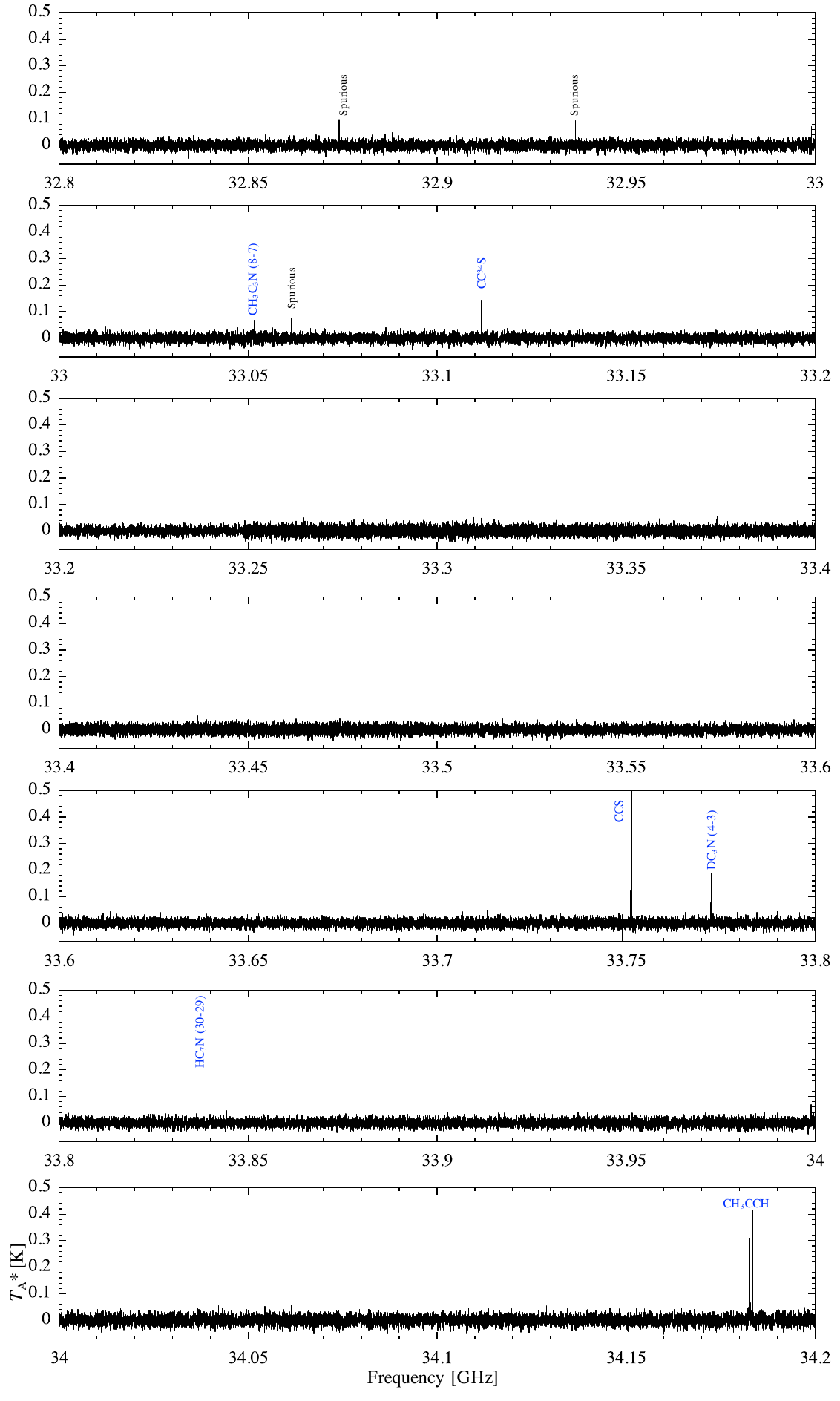} 
 \end{center}
\caption{Close-up spectra (Continued). {Alt text: Seven spectral plots. The vertical axis shows the antenna temperature in the unit of Kelvin, and the horizontal axis shows the frequency in the unit of gigahertz.}}\label{fig:s3}
\end{figure*}

\begin{figure*}
\setcounter{figure}{7}
 \begin{center}
  \includegraphics[bb= 0 15 600 900, scale = 0.72]{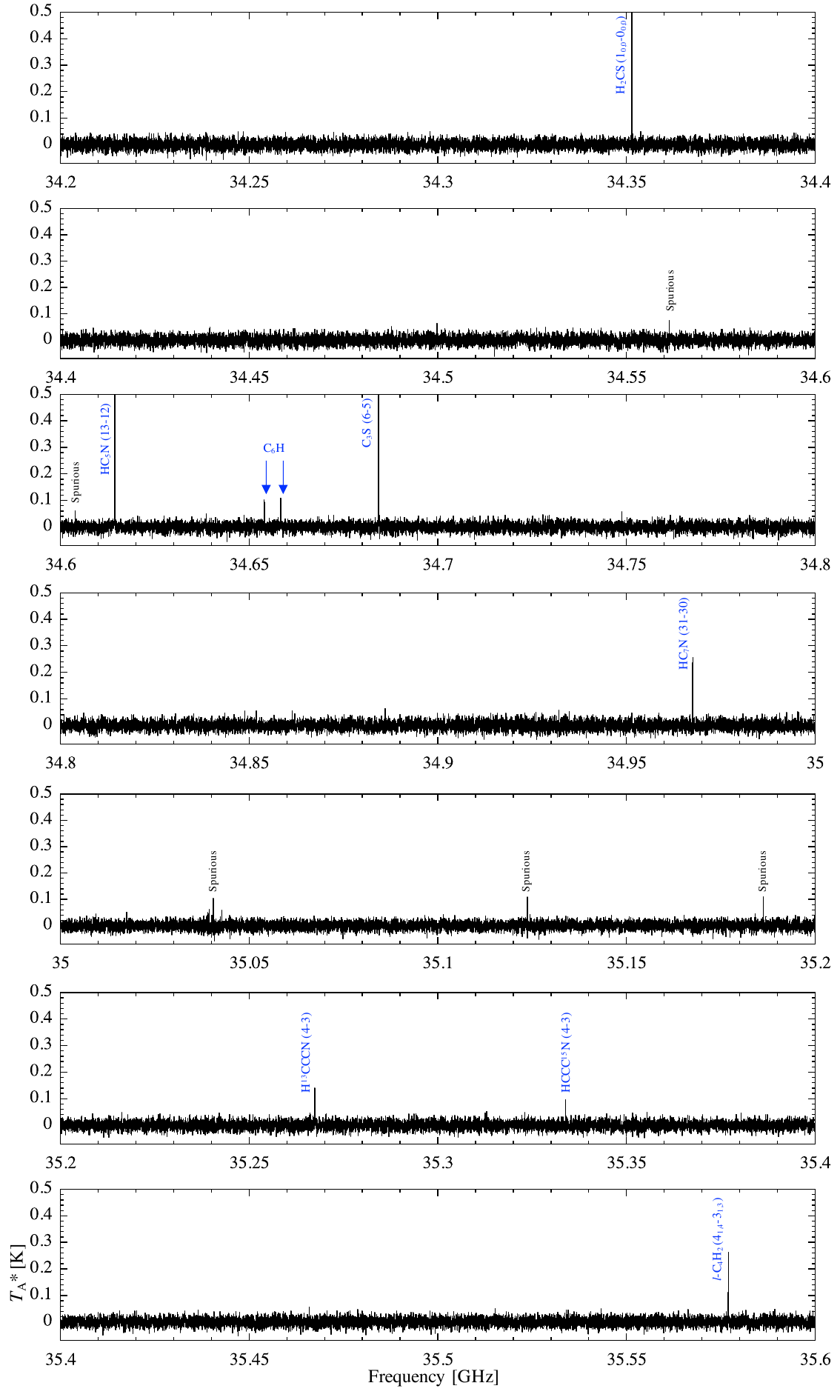} 
 \end{center}
\caption{Close-up spectra (Continued). {Alt text: Seven spectral plots. The vertical axis shows the antenna temperature in the unit of Kelvin, and the horizontal axis shows the frequency in the unit of gigahertz.}}\label{fig:s4}
\end{figure*}

\begin{figure*}
\setcounter{figure}{7}
 \begin{center}
  \includegraphics[bb= 0 15 600 900, scale = 0.72]{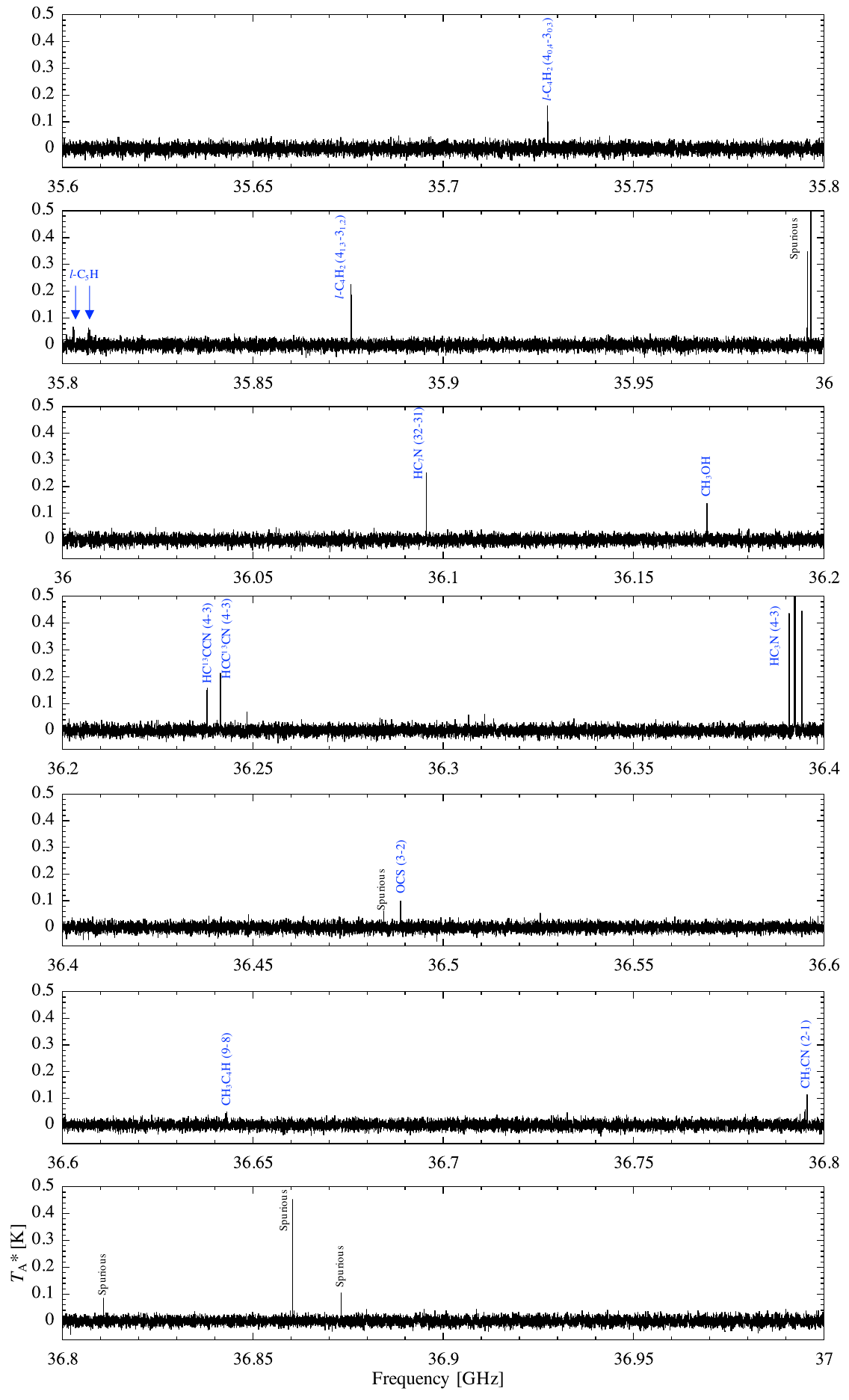} 
 \end{center}
\caption{Close-up spectra (Continued). {Alt text: Seven spectral plots. The vertical axis shows the antenna temperature in the unit of Kelvin, and the horizontal axis shows the frequency in the unit of gigahertz.}}\label{fig:s5}
\end{figure*}

\begin{figure*}
\setcounter{figure}{7}
 \begin{center}
  \includegraphics[bb= 0 15 600 900, scale = 0.72]{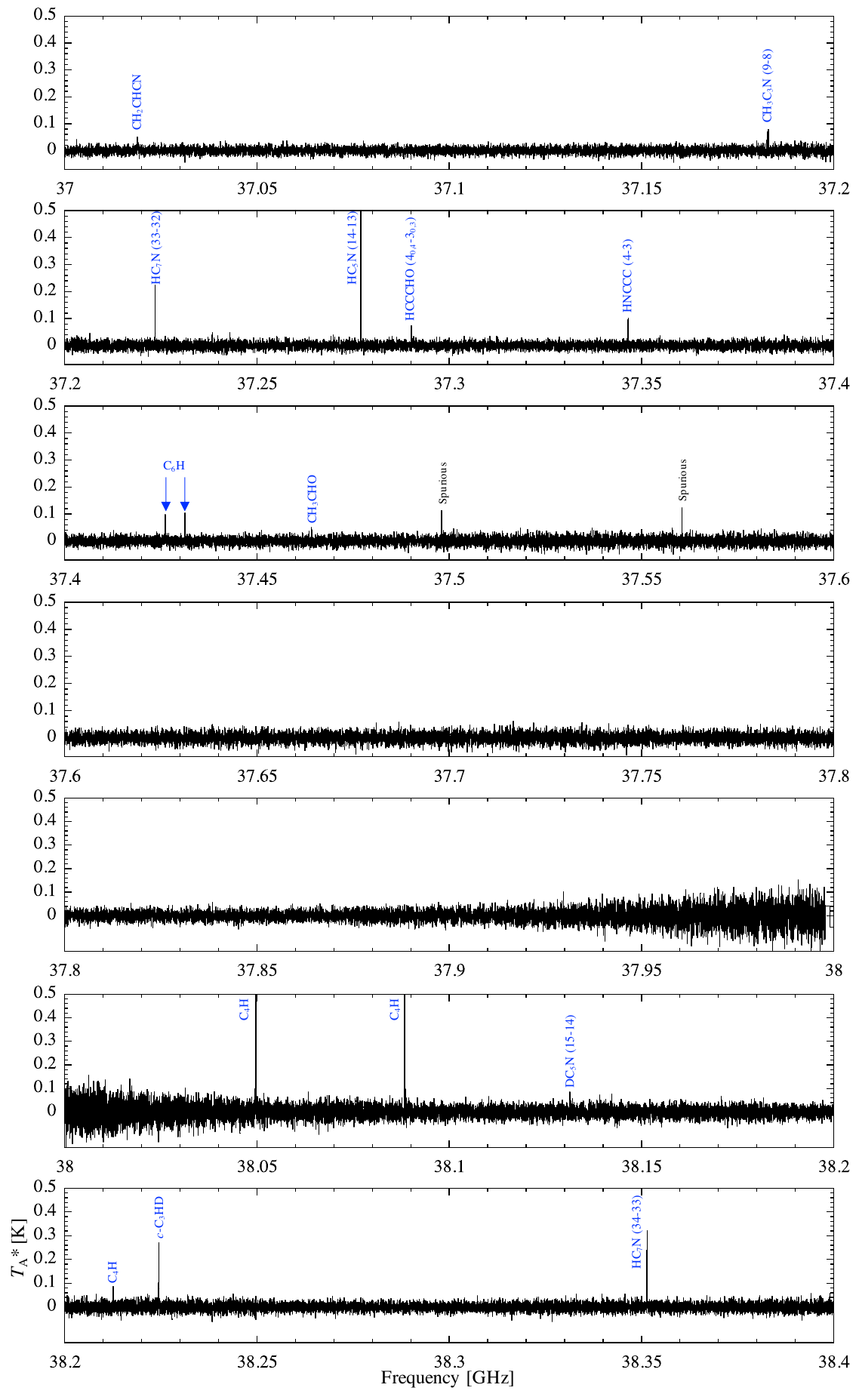} 
 \end{center}
\caption{Close-up spectra (Continued). {Alt text: Seven spectral plots. The vertical axis shows the antenna temperature in the unit of Kelvin, and the horizontal axis shows the frequency in the unit of gigahertz.}}\label{fig:s6}
\end{figure*}

\begin{figure*}
\setcounter{figure}{7}
 \begin{center}
  \includegraphics[bb= 0 15 600 900, scale = 0.72]{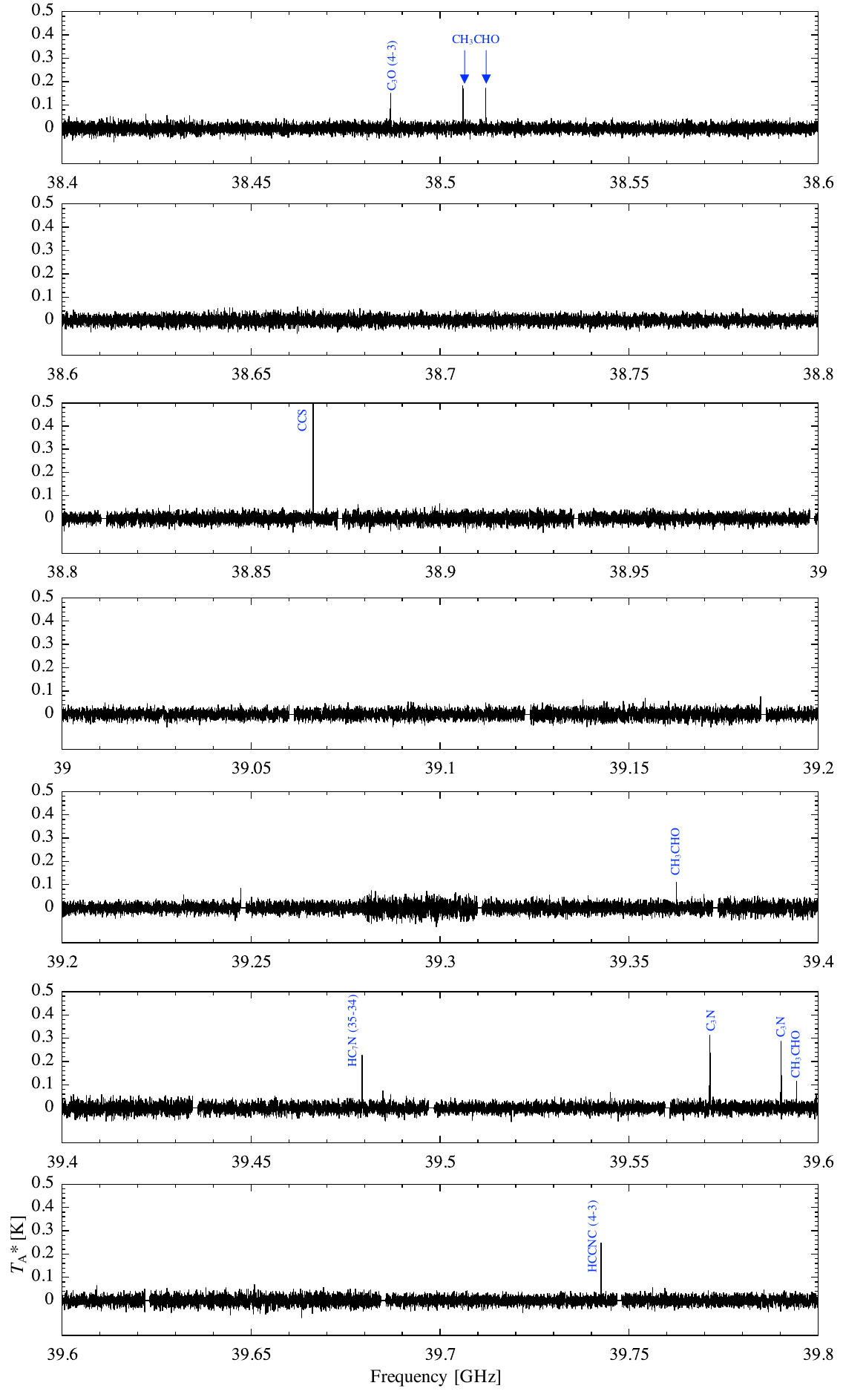} 
 \end{center}
\caption{Close-up spectra (Continued). {Alt text: Seven spectral plots. The vertical axis shows the antenna temperature in the unit of Kelvin, and the horizontal axis shows the frequency in the unit of gigahertz.}}\label{fig:s7}
\end{figure*}

\begin{figure*}
\setcounter{figure}{7}
 \begin{center}
  \includegraphics[bb= 0 15 600 900, scale = 0.72]{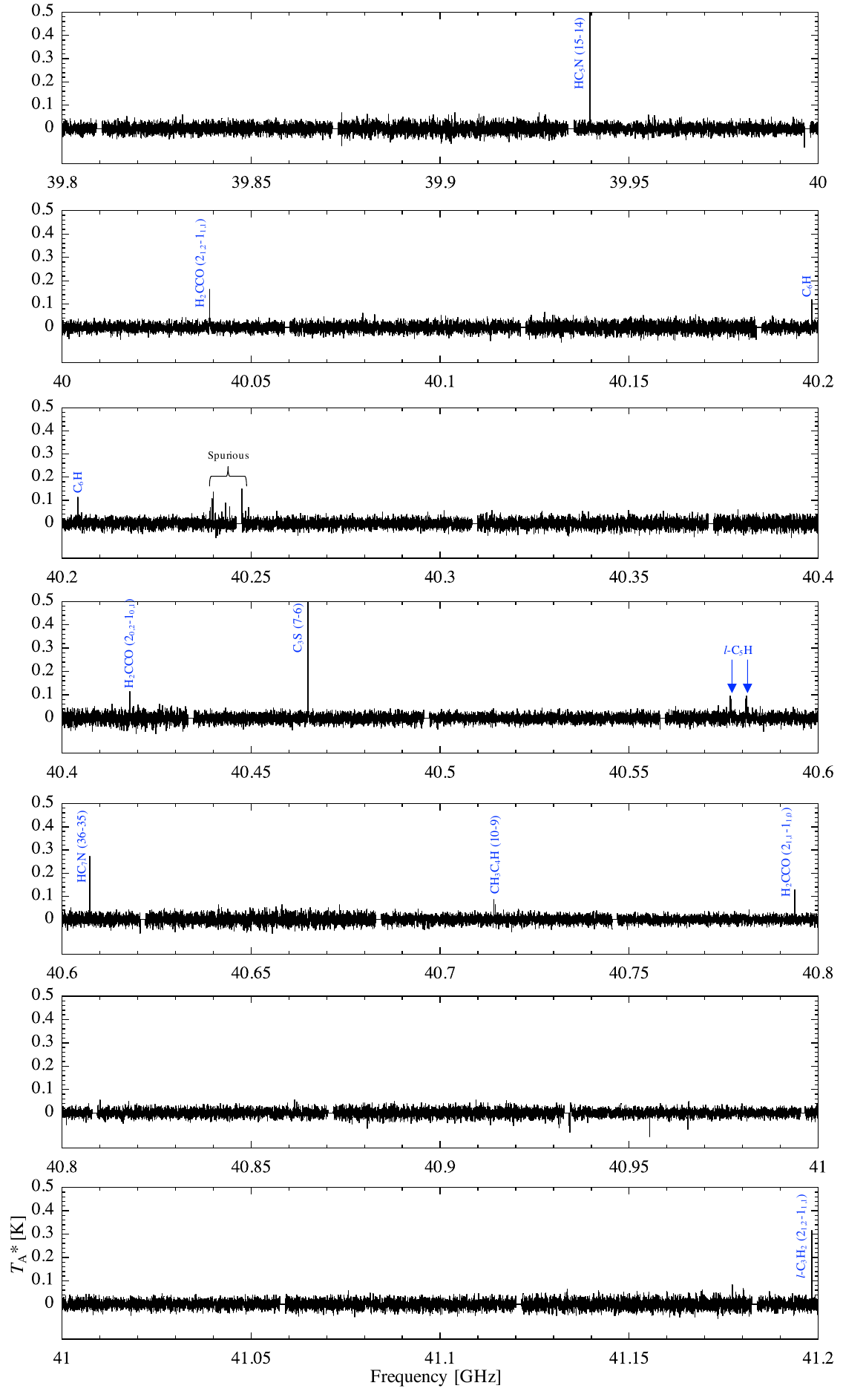} 
 \end{center}
\caption{Close-up spectra (Continued). {Alt text: Seven spectral plots. The vertical axis shows the antenna temperature in the unit of Kelvin, and the horizontal axis shows the frequency in the unit of gigahertz.}}\label{fig:s8}
\end{figure*}

\begin{figure*}
\setcounter{figure}{7}
 \begin{center}
  \includegraphics[bb= 0 15 600 900, scale = 0.72]{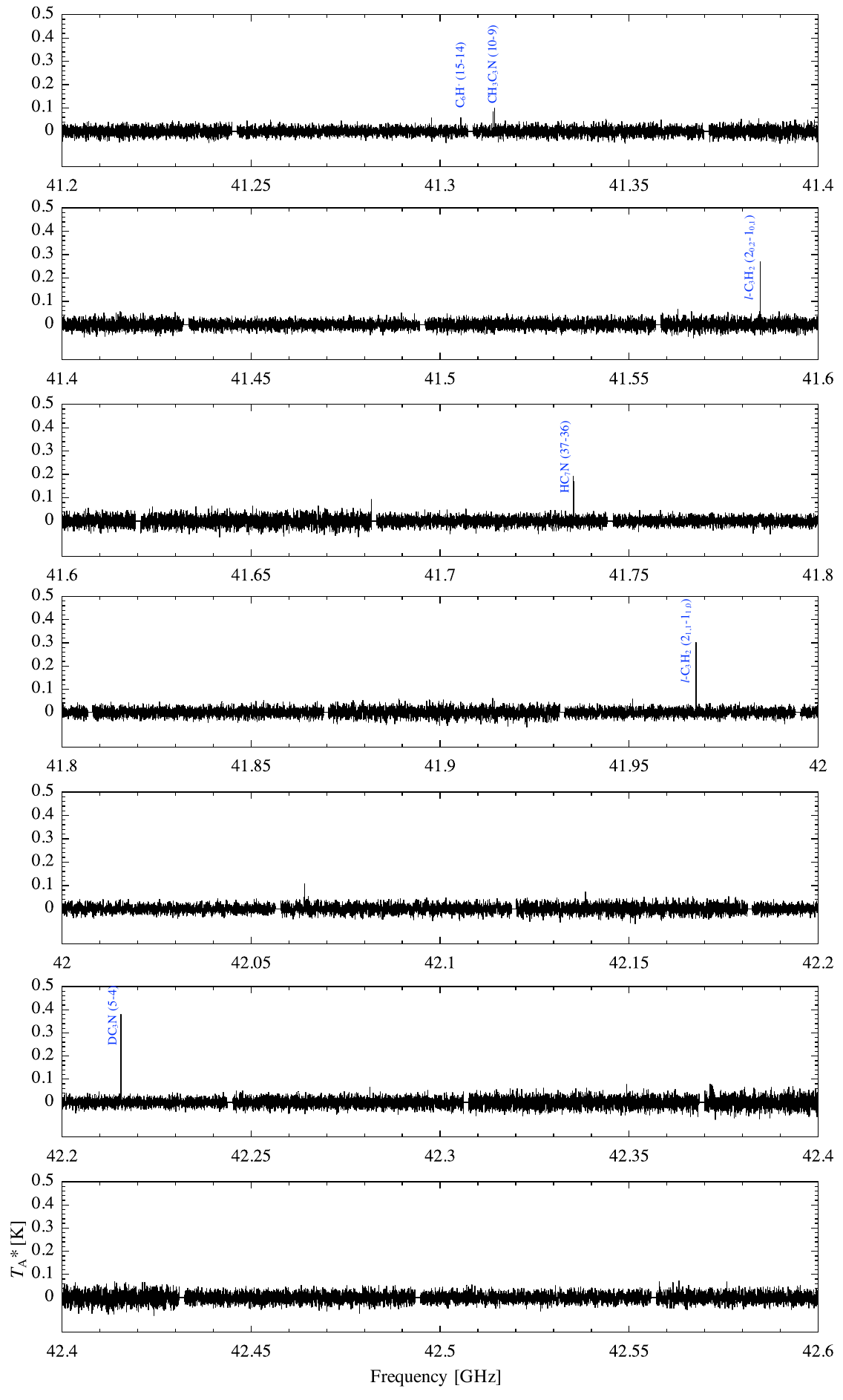} 
 \end{center}
\caption{Close-up spectra (Continued). {Alt text: Seven spectral plots. The vertical axis shows the antenna temperature in the unit of Kelvin, and the horizontal axis shows the frequency in the unit of gigahertz.}}\label{fig:s9}
\end{figure*}

\begin{figure*}
\setcounter{figure}{7}
 \begin{center}
  \includegraphics[bb= 0 15 600 900, scale = 0.72]{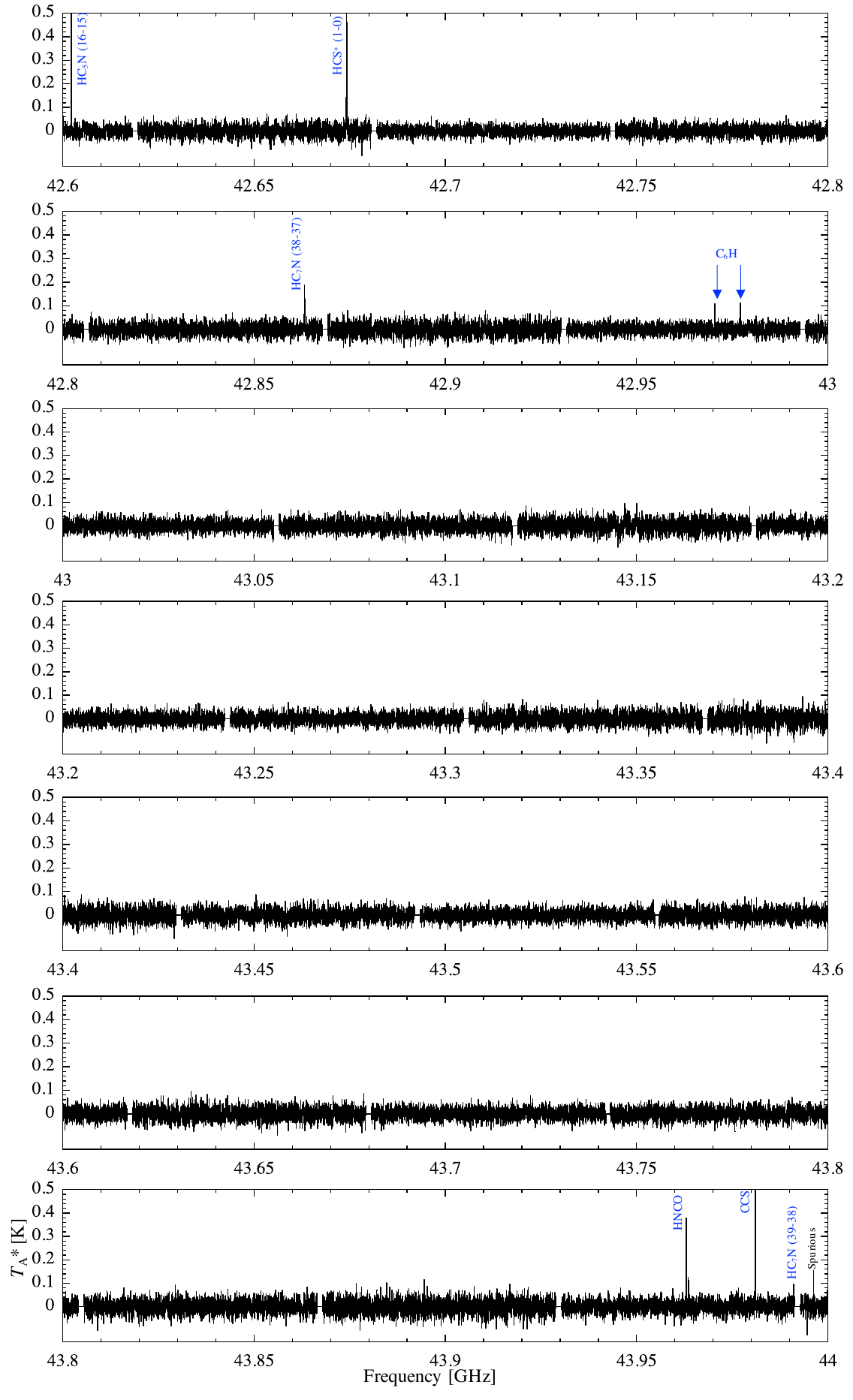} 
 \end{center}
\caption{Close-up spectra (Continued). {Alt text: Seven spectral plots. The vertical axis shows the antenna temperature in the unit of Kelvin, and the horizontal axis shows the frequency in the unit of gigahertz.}}\label{fig:s10}
\end{figure*}

\begin{figure*}
\setcounter{figure}{7}
 \begin{center}
  \includegraphics[bb= 0 15 600 900, scale = 0.72]{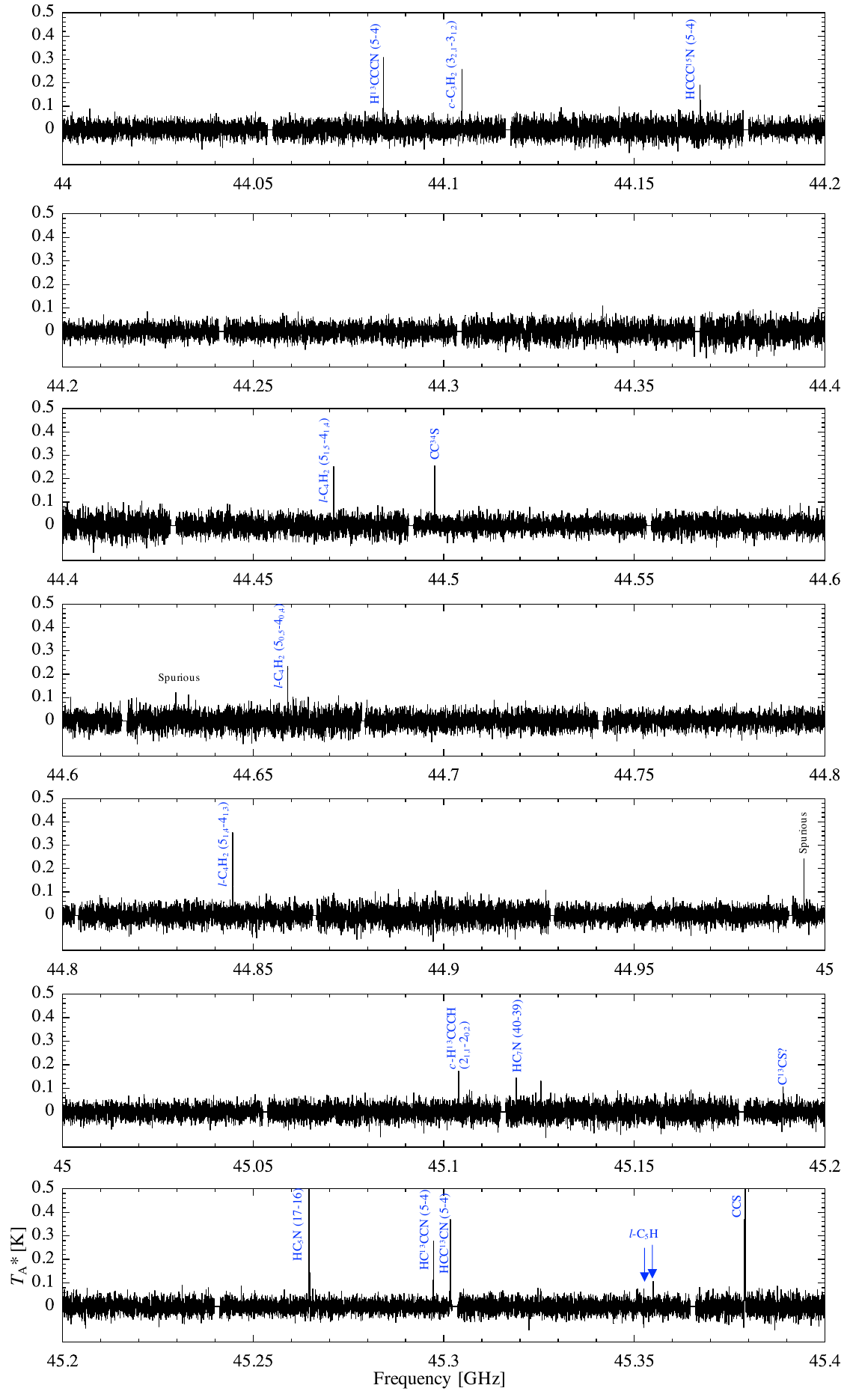} 
 \end{center}
\caption{Close-up spectra (Continued). {Alt text: Seven spectral plots. The vertical axis shows the antenna temperature in the unit of Kelvin, and the horizontal axis shows the frequency in the unit of gigahertz.}}\label{fig:s11}
\end{figure*}

\begin{figure*}
\setcounter{figure}{7}
 \begin{center}
  \includegraphics[bb= 0 15 600 900, scale = 0.72]{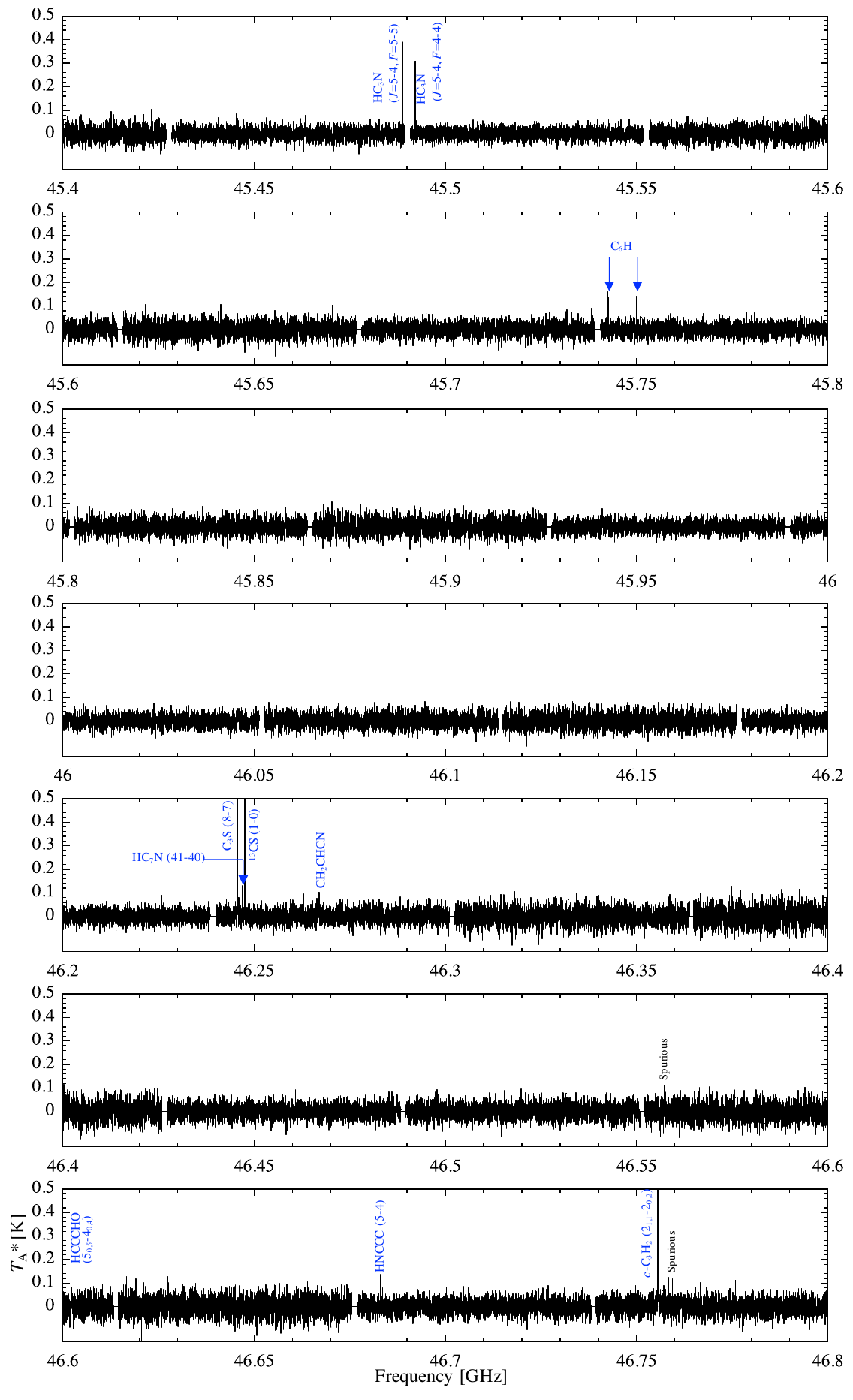} 
 \end{center}
\caption{Close-up spectra (Continued). {Alt text: Seven spectral plots. The vertical axis shows the antenna temperature in the unit of Kelvin, and the horizontal axis shows the frequency in the unit of gigahertz.}}\label{fig:s12}
\end{figure*}

\begin{figure*}
\setcounter{figure}{7}
 \begin{center}
  \includegraphics[bb= 0 15 600 900, scale = 0.72]{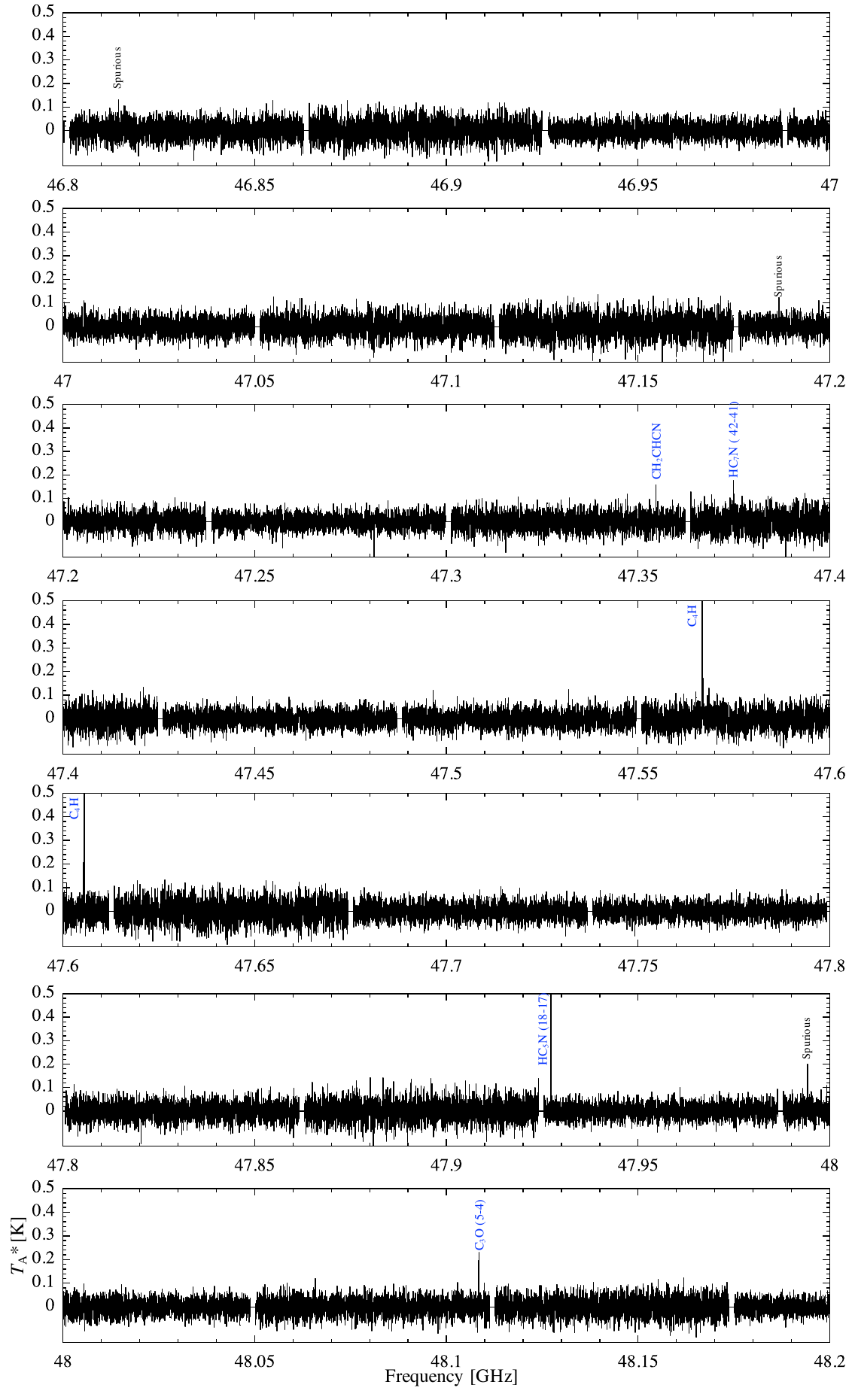} 
 \end{center}
\caption{Close-up spectra (Continued). {Alt text: Seven spectral plots. The vertical axis shows the antenna temperature in the unit of Kelvin, and the horizontal axis shows the frequency in the unit of gigahertz.}}\label{fig:s13}
\end{figure*}

\begin{figure*}
\setcounter{figure}{7}
 \begin{center}
  \includegraphics[bb= 0 15 600 900, scale = 0.72]{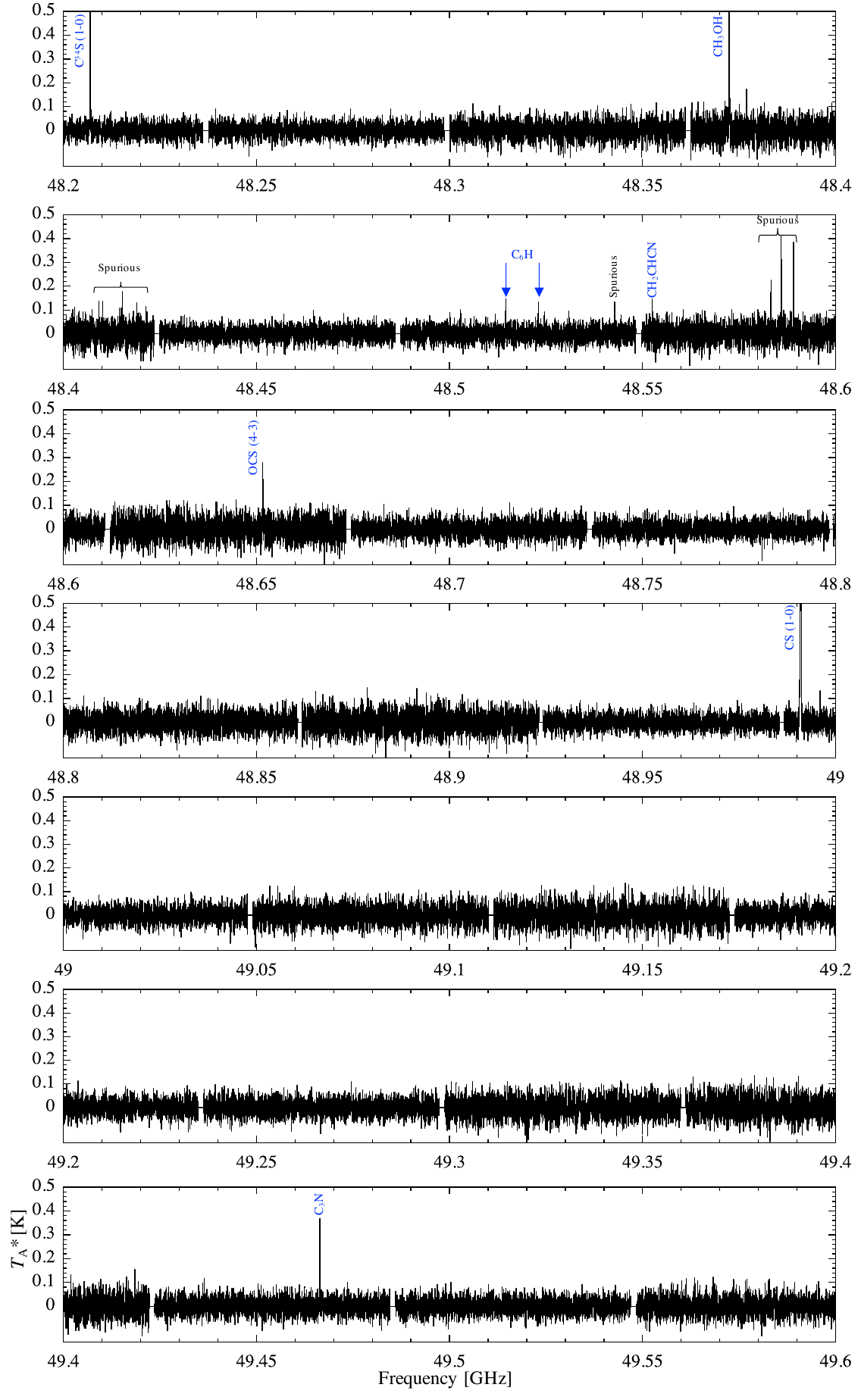} 
 \end{center}
\caption{Close-up spectra (Continued). {Alt text: Seven spectral plots. The vertical axis shows the antenna temperature in the unit of Kelvin, and the horizontal axis shows the frequency in the unit of gigahertz.}}\label{fig:s14}
\end{figure*}

\begin{figure*}
\setcounter{figure}{7}
 \begin{center}
  \includegraphics[bb= 0 15 600 280, scale = 0.72]{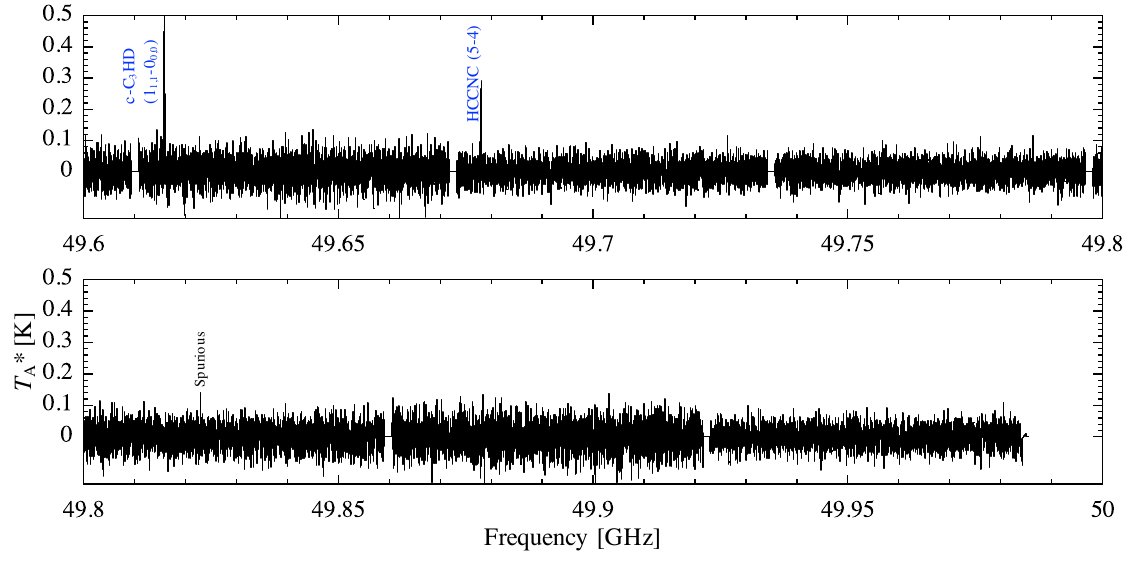} 
 \end{center}
\caption{Close-up spectra (Continued). {Alt text: Two spectral plots. The vertical axis shows the antenna temperature in the unit of Kelvin, and the horizontal axis shows the frequency in the unit of gigahertz.}}\label{fig:s15}
\end{figure*}

\begin{figure*}
 \begin{center}
  \includegraphics[bb= 0 15 800 190, width = \textwidth]{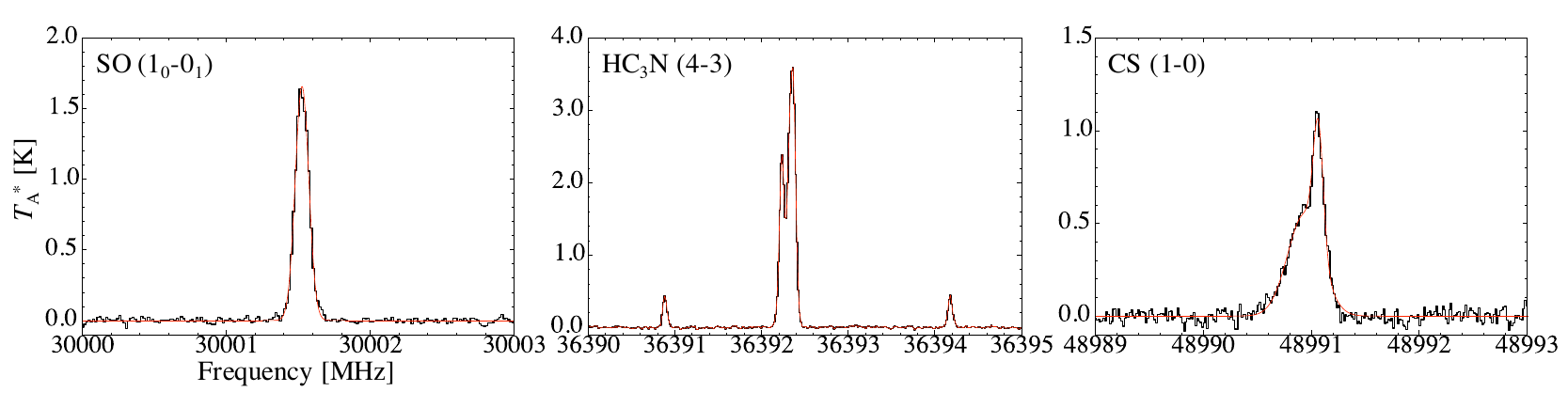} 
 \end{center}
\caption{Spectra of SO ($J_N=1_0-0_1$), HC$_3$N ($J=4-3$), and CS (1--0). The black lines and red curves indicate the observed spectra and the Gaussian fitting results, respectively. {Alt text: Three spectral plots. The vertical axis shows the antenna temperature in the unit of Kelvin, and the horizontal axis shows the frequency in the unit of megahertz.}}\label{fig:3spec}
\end{figure*}


\end{document}